\documentclass[preprint,10pt]{elsarticle} 
\usepackage{graphicx,subcaption}
\usepackage{natbib}
\usepackage{colortbl}
\usepackage{amssymb}
\usepackage{moreverb} % JAHO
 \usepackage{float}   % JAHO
\usepackage{listings}
\usepackage[linesnumbered,ruled]{algorithm2e}
\SetCommentSty{mycommfont}
\usepackage{nomencl}
\usepackage[english]{babel}
\usepackage{nomencl}  % it doesn't compile !!!! See MISCELL.doc 
\usepackage{leftidx} % For attaching left-hand sided scripts
 \usepackage{amsmath, amsthm, amssymb}

\usepackage[left=1.5cm,top=2cm,right=1.5cm,bottom=2.5cm]{geometry} % 
\usepackage{color}
\usepackage{listings}
\usepackage{mathrsfs}

\newtheorem{remark}{Remark}[section]

\definecolor{mygreen}{rgb}{0,0.6,0}
\definecolor{mygray}{rgb}{0.5,0.5,0.5}
\definecolor{mymauve}{rgb}{0.58,0,0.82}
\usepackage{hyperref}
\usepackage{algorithmic}
\usepackage{tikz}
\usepackage{physics}
\usepackage{framed}
\usepackage{multicol}
\usepackage{multirow}
\newcommand\algorithmicprocedure{\textbf{procedure}}
\newcommand{\algorithmicendprocedure}{\algorithmicend\ \algorithmicprocedure}
\makeatletter
\newcommand\PROCEDURE[3][default]{%
  \ALC@it
  \algorithmicprocedure\ \textsc{#2}(#3)%
  \ALC@com{#1}%
  \begin{ALC@prc}%
}
\newcommand\ENDPROCEDURE{%
  \end{ALC@prc}%
  \ifthenelse{\boolean{ALC@noend}}{}{%
    \ALC@it\algorithmicendprocedure
  }%
}
\newenvironment{ALC@prc}{\begin{ALC@g}}{\end{ALC@g}}
\makeatother
\usepackage{bbm}
\makenomenclature
\usepackage[most]{tcolorbox}
\usepackage{xcolor}
\usepackage{colortbl}
\definecolor{mydarkgreen}{RGB}{0,100,0}
\colorlet{customgreen}{green!50!mydarkgreen}

\newcommand{\Deltaq}{{\Delta \hat{u}}}
\newcommand{\uref}{{\mathbf{u}_{\text{ref}}}}
\newcommand{\rtle}{{\mathbf{R}_t^{Le}}}
\newcommand{\barrte}{{\Bar{\mathbf{R}}_t^e}}
\newcommand{\utk}{{(\mathbf{u}_{t}^{(k)};\boldsymbol{\mu})}}
\newcommand{\tildeutkm}{{(\mathbf{\Tilde{u}}_t^{(k)};\boldsymbol{\mu})}}
\newcommand{\tildeutk}{{\mathbf{\Tilde{u}}_t^{(k)}}}
\newcommand{\tildeu}{{\mathbf{\Tilde{u}}}}
\newcommand{\ut}{{(\mathbf{u}_{t};\boldsymbol{\mu})}}
\newcommand{\deltauhat}{{\mathbf{\Delta \hat{u}}}}
\newcommand{\tildeutm}{{(\mathbf{\Tilde{u}}_t;\boldsymbol{\mu})}}
\newcommand{\deltatildeu}{{\mathbf{\Delta \Tilde{u}}}}
\newcommand{\hatptk}{{\mathbf{\hat{p}}_t^{(k)}}}
\newcommand{\hatpt}{{\mathbf{\hat{p}}_t}}
\begin{document}

\begin{frontmatter}

\title{Hyper-reduction for Petrov-Galerkin reduced order models.}

\author[1,2]{S. Ares de Parga\corref{cor1}}
\ead{sebastian.ares@upc.edu}
\author[1,2]{J.R. Bravo}
\author[1,2,3]{J.A. Hern\'{a}ndez}
\author[1,2]{R. Zorrilla}
\author[1,2]{R. Rossi}

\cortext[cor1]{Corresponding author}

\address[1]{Universitat Polit\`{e}cnica de Catalunya, Department of Civil and Environmental Engineering, Building B0, Campus Nord, Jordi Girona 1-3, Barcelona 08034, Spain}
\address[2]{Centre Internacional de M\`{e}todes Num\`{e}rics en Enginyeria (CIMNE), Universitat Polit\`{e}cnica de Catalunya, Building C1, Campus Nord, Jordi Girona 1-3, Barcelona 08034, Spain}%
\address[3]{Universitat Polit\`{e}cnica de Catalunya, E.S. d'Enginyeries Industrial, Aeroespacial i Audiovisual de Terrassa, C/ Colom, 11,  Terrassa 08222, Spain}%

 \begin{abstract}
Projection-based Reduced Order Models are based on the idea of minimizing the discrete residual of a ``full order model" (FOM) while at the same time constraining the unknowns to live in a space of reduced dimension. For problems with symmetric positive definite (SPD) Jacobians, this minimization can be achieved optimally by projecting the full order residual onto the approximation basis (Galerkin Projection). This approach is sub-optimal for problems with non-SPD Jacobians since it only guarantees that the projection of the residual onto the chosen basis is minimized and not the residual itself. One possible alternative in such cases is to directly minimize the 2-norm of the residual. This minimization can be achieved either by using QR factorization to solve the resulting least-squares problem or by employing the method of the normal equations (LSPG) to the same end. The first approach involves constructing and factorizing a rectangular tall and skinny matrix of size proportional to the number of unknowns of the FOM. The LSPG method avoids the use of the large matrix by directly constructing the product by its transpose. Unfortunately, constructing such a product element by element is not feasible and requires the use of a complementary mesh, which adds an extra layer of complexity to the hyper-reduction process when performing mesh sampling.
The main idea of this work is to propose an alternative technique based on the idea of Petrov-Galerkin minimization. Essentially, we choose a left basis so that a least-squares minimization procedure can be carried out on a reduced problem while guaranteeing that the discrete full order residual is minimized. The resulting procedure is applicable to problems with both SPD and non-SPD Jacobians. Additionally, the resulting minimization problem can be assembled element by element, avoiding the use of the complementary mesh and simplifying implementation in the context of finite elements. The resulting technique is amenable to hyper-reduction by the use of the Empirical Cubature Method and can be readily applied in the context of nonlinear reduction procedures.

\end{abstract}


\end{frontmatter}
\noindent
\textbf{Note:} This is the accepted manuscript version of the article published in Computer Methods in Applied Mechanics and Engineering. The final, definitive version is available at \href{https://doi.org/10.1016/j.cma.2023.116298}{https://doi.org/10.1016/j.cma.2023.116298}. This manuscript is distributed under the CC-BY-NC-ND license.
\section{Introduction}
Projection-based reduced order models (PROM) can significantly reduce the time and storage required for the solution of high-fidelity (high-dimensional) discretized PDE based computational models for a given range of parametric settings, also known as full-order models (FOM) in the literature. These PROMs are intended for the creation of time-critical models such as digital twins, design/shape optimization, control problem optimization, and other applications. 
Unfortunately, the cost of creating a PROM of dimension $n$ scales with both $n$ and the dimension of the underlying FOM $N>>n$. To overcome this problem, one popular approach is to divide the computation into two parts: one that scales with the FOM but can be executed offline to compute some numerical quantities; and another that uses the aforementioned numerical quantities to perform all online computations with a complexity that is independent of the FOM's large dimension. 
The first part of this offline-online decomposition is at the heart of model order reduction, where the state of a large-scale system resides on a lower dimensional manifold engendered by time evolution and/or input parameter variation. To do this, solution snapshots corresponding to a carefully chosen set of input parameters are subjected to dimensionality reduction (compression) in order to determine a reduced-order basis (ROB), using, for instance, Proper Orthogonal Decomposition (POD) \cite{Sirovich1987, Balachandar1998}. The FOM is then projected onto the subspace spanned by this ROB in order to provide rapid and accurate online numerical predictions. This offline-online technique has been shown to be effective for parametric, linear problems \cite{Antoulas2005, Cuong05}, as well as nonlinear problems with a low-order polynomial dependency \cite{Rowley2004,Benner2015}. 
\\
\indent Similarly, computational methods known as hyper-reduction have gained popularity in overcoming computational bottlenecks caused by repeated re-evaluations of parametric reduced-order operators. These methods approximate these operators with a computational cost that is independent of the FOM, trading some of the PROM's accuracy for speed. Carlberg first introduced the classification of hyper-reduction methods into two types: the approximate-then-project and the project-then-approximate, as described in \cite{Carlberg2011}. The approximate-then-project hyper-reduction methods approximate first an operator of interest and then project the approximation onto the left ROB. The underlying idea for avoiding a computational cost that grows with the problem's large dimension $N$ may be traced back to the gappy POD technique, which was initially devised for image reconstruction \cite{Everson1995}. After nearly a decade, the empirical interpolation method was launched on a continuous level with its discrete variant known as discrete EIM (DEIM) \cite{Chaturantabut2010}, which is likely the most popular approximate-then-project hyper-reduction method to date. Other approximate-then-project hyper-reduction methods have been introduced, such as the missing point estimation \cite{Galbally2010} and the collocation method \cite{Ryckelynck2005}. Another noteworthy approximate-then-project hyper-reduction method is the Gauss-Newton with approximated tensors (GNAT) method \cite{Carlberg2013}, which was conceived for the Petrov-Galerkin projection rather than the Galerkin framework. \
On the other hand, hyper-reduction methods of the project-then-approximate type approximate directly the projection onto the left ROB. They can be regarded as extended quadrature rules, in which the set of quadrature ``points" and associated weights are learned via a supervised method on an empirical set of training data. Among this family of methods, we can find the energy-conserving sampling and weighting (ECSW) method developed in Ref. \cite{Farhat2015}, and the empirical cubature method (ECM) \cite{Joaquin2016,Joaquin2020}. These methods compute a subset of the underlying FOM's elements (or other entities) that define the quadrature points, then construct an approximation of the full solution, which is commonly known as mesh sampling. \\ \\
Hyper-reduction methods have been widely demonstrated to produce numerically stable hyper-reduced PROMs (HPROM) \cite{Chaturantabut2010, Farhat2015, Galbally2010, Joaquin2016, Joaquin2020, Ryckelynck2005, Barrault2004, Farhat2014}, especially for standard Galerkin PROM. By definition, in a Galerkin projection, the left ($\Psi$) and right ($\Phi$) ROBs are set to be equal, or the test and approximation spaces are the same ($\Psi=\Phi$). The first methods that explored the acceleration of Petrov-Galerkin (PG) PROM, where the left and right ROBs differ ($\Psi\neq\Phi$), are of the approximate-then-project type. These include a gappy-POD-like method, similar to DEIM, called the Gauss-Newton with approximated tensors (GNAT) method \cite{Carlberg2011, Amsallem2012, Carlberg2013, Carlberg2017}, and a least-squares variant of the collocation method \cite{Ryckelynck2005}. The GNAT was developed in the context of PG-PROMs, especially for PG-PROMs in which the left ROB is chosen to minimize the discrete, nonlinear residual over the approximation subspace associated with the right ROB in the 2-norm. In this scenario, applying Newton's technique to solve the system of equations is equal to using the Gauss-Newton method to solve the nonlinear, least-squares minimization problem. As a result, the PG-PROM method on which GNAT is based is commonly referred to as the least-squares Petrov-Galerkin (LSPG) projection method in the literature \cite{Carlberg2011}. Essentially, when the Jacobian ($\mathbf{J}$) associated with the Gauss-Newton method results in a symmetric positive definite (SPD) operator, the Galerkin PROM can be shown to minimize the discrete, nonlinear residual in a $\mathbf{J}^{-1}$-norm, whereas the LSPG-PROM can be used for the general case where the Jacobian associated with the Gauss-Newton method is not SPD. For example, when the Jacobian comes from the discretization of the Navier-Stokes equations, the Galerkin PROM approach lacks the minimum-residual optimality property. However, all approximate-then-project methods, such as GNAT, are based on sub-optimal greedy algorithms that take the size of the reduced mesh as input, which is unknown a priori.
Recently, a project-then-approximate method for the hyper-reduction of PG-PROMs was developed, presenting an Energy-Conserving Sampling and Weighting (ECSW) type method \cite{Farhat2015}. This method was extended to PROMs based on local, piecewise-affine approximation subspaces and generalized to finite elements spatial discretizations, as well as finite volume, and finite differences semi-discretizations. \\
However, in order to evaluate the hyper-reduced residual vectors and Jacobian matrices associated with the selected elements and their neighbors, PG-PROMs require a complementary mesh to be constructed \cite{Grimberg2021}. This means that the hyper-reduced sample mesh must include the patch of elements that contains the selected elements as a complementary mesh, significantly increasing the number of elements required to integrate the hyper-reduced model \footnote{This behavior should not be confused with hyper-reducing a finite volume model \cite{Carlberg2013,Blonigan2020,Shimizu2021}, where it is shown that for the collocation mesh, it is necessary to include neighboring cells (and neighbors of neighbors for second order schemes) to compute the governing equations.}. The issue arises from the iteration-dependent left ROB of the LSPG-PROM. \\ \\
In this paper, we propose a novel approach to address the problem of using a complementary mesh for hyper-reduction by introducing an equivalent invariant left ROB $\mathbf{\Psi}$ that preserves the 2-norm minimum-residual optimality. We obtain two different invariant ROBs ($\mathbf{\Psi}\neq\mathbf{\Phi}$) and an alternative PG-PROM by compressing either the converged projected Jacobians onto the right ROB or the non-converged residuals.  Additionally, we utilize the empirical cubature method (ECM) to carry out the PG-PROM hyper-reduction without the need for a complementary mesh. ECM identifies a reduced subset of elements and corresponding positive weights, which are calculated from a minimization of the entire unassembled residual projected onto the left ROB $\mathbf{\Psi}$. Our proposed approach offers a novel solution to the problem of avoiding the use of a complementary mesh for hyper-reduction and represents a significant contribution to the field.
\section{Parametrized Problem}
\label{Parametrized Problem}
Let us consider a general nonlinear parametrized discrete operator of the form
\begin{equation}
\mathbf{R}_{t}\ut = \mathbf{R}(\uref + \mathbf{\Delta u}(t;\boldsymbol{\mu}), t;\boldsymbol{\mu}) = \mathbf{0},
\label{residual_FOM}
\end{equation}
where $t \in [0, T]$ represents time up to a final time $T$. With a minor abuse of notation, we use the subscript $t$ to denote that a variable is evaluated at a specific time instance $t$. The term $\mathbf{u}_t=\uref + \mathbf{\Delta u}_t \in \mathbb{R}^N$ designates the nodal state variable evaluated at the specific time instance $t$. Here, $\uref$ represents a reference solution, often based on the solution at the previous time step, $\mathbf{u}_{t-1}$.\footnote{If necessary, the reference solution $\uref$ can be enhanced with methods like extrapolation based on the derivatives of $\mathbf{u}_{t-1}$ with respect to time. This can provide more accurate initial conditions, aiding in the stability and convergence of the iterative scheme.} We have intentionally omitted the explicit dependency of $\mathbf{u}_t$ on the parameter $\boldsymbol{\mu}$ for simplicity of exposition. However, it is important to emphasize that $\mathbf{u}_t$ intrinsically depends on $\boldsymbol{\mu}$, and this dependency will be explicitly denoted when necessary for clarity and instructional purposes. The vector $\boldsymbol{\mu}$ $\in$ $\mathcal{D}$ denotes a specific instance of the input parameters, where $\mathcal{D} \subset \mathbb{R}^P$ represents the discrete point-set of all possible input parameter vectors. The discrete operator $\mathbf{R}_t:\mathbb{R}^N\times \mathcal{D}\rightarrow \mathbb{R}^N$ will be referred to as the residual vector. For simplicity of exposition, homogeneous Dirichlet boundary conditions are considered. This high fidelity model will be referred to as the full-order model (FOM). The residual vector is assembled in the usual manner by summing the elemental contributions, which can be expressed as:
\begin{equation}
    \mathbf{R}_t=\sum_{e=1}^{L}\mathbf{L}^{e^T}\mathbf{R}_t^e,
    \label{FOM Assembly}
\end{equation}
where $L$ denotes the total number of elements, and $\mathbf{L}^e$ is the assembly operator that connects the degrees of freedom (DOFs) of element $e$ with the global DOFs. The vector $\mathbf{R}_t^e$ represents the values of the elemental residual associated with the DOFs of element $e$.\\
\\
An iterative scheme, such as the Newton-Raphson strategy, is required to find a solution $\mathbf{u}_t$ for the nonlinear residual $\mathbf{R}_t$, resulting in the following iterations: for $k=1,\hdots,K$, solve
\begin{equation}
\boxed{\begin{aligned}
\mathbf{J}_{t}^{(k)}\utk \mathbf{p}_{t}^{(k)} & = -\mathbf{R}_{t}^{(k)}\utk\\
\mathbf{\Delta u}_{t}^{(k+1)} & =\mathbf{\Delta u}_{t}^{(k)}+\alpha_{t}^{(k)}\mathbf{p}_{t}^{(k)}\\
\mathbf{u}_{t}^{(k+1)}&=\mathbf{u}_{t}^{(k)}+\mathbf{\Delta u}_{t}^{(k+1)},
\end{aligned}}
\label{Newton-Raphson FOM}
\end{equation}
where $\mathbf{J}$ represents the Jacobian operator, and $\mathbf{\Delta u}$ is the increment of the solution state variable at each iteration. $\mathbf{R}_t^{(k)}$ and $\mathbf{J}_t^{(k)}$ are the residual and Jacobian evaluated at the current iterative solution $\mathbf{u}_{t}^{(k)}$. Specifically, $\mathbf{R}_t^{(k)}=\left.\mathbf{R}_t\right|_{\mathbf{u}_t^{(k)}} = \mathbf{R}(\uref+\mathbf{\Delta u}_{t}^{(k)},t;\boldsymbol{\mu})$, and $\mathbf{J}_t^{(k)}=\left.\frac{\partial \mathbf{R}_t}{\partial \mathbf{u}}\right|_{\mathbf{u}_t^{(k)}}=\mathbf{J}(\uref+\mathbf{\Delta u}_{t}^{(k)},t;\boldsymbol{\mu})$.

The process carries out iterations until convergence is reached, thereby finding the solution $\mathbf{u}_t$ for the specified time step $t$ and the given parameter set $\boldsymbol{\mu}$. Notably, a line search could potentially be employed in the direction of $\mathbf{p}\in \mathbb{R}^N$ to determine the step length $\alpha$. For a comprehensive understanding and more specific details about the procedure, please refer to Algorithm \ref{alg:Newton FOM}. We used the Newton-Raphson method throughout this work for clarity of presentation. However, it is worth noting that if the Jacobian matrix is not available or computationally expensive to compute, one could use a fixed-point iteration method such as Picard's method instead of the Newton-Raphson method.
\begin{algorithm}
\caption{Newton-Raphson strategy for FOM.}
\label{alg:Newton FOM}
\hspace*{\algorithmicindent} \textbf{Input:} State variable $\uref$\\
\hspace*{\algorithmicindent} \textbf{Output:} State variable $\mathbf{u}_t$ and increment update $\mathbf{\Delta u}_t$
\begin{algorithmic}[1]
    \STATE Initialize $\mathbf{\Delta u}_t^{(0)}=\mathbf{0}$, i.e., $\mathbf{u}_t^{(0)}=\uref$
    \FOR{$k=0,\hdots,K $ (convergence)}
        \STATE Evaluate residual $\mathbf{R}_t^{(k)}=\left.\mathbf{R}_t\right|_{\mathbf{u}_t^{(k)}}$
        \STATE Evaluate Jacobian $\mathbf{J}_t^{(k)}=\left.\frac{\partial \mathbf{R}_t}{\partial \mathbf{u}}\right|_{\mathbf{u}_t^{(k)}}$
        \STATE Solve $\mathbf{J}_{t}^{(k)}  \mathbf{p}_t^{(k)} = -\mathbf{R}_t^{(k)}$
        % \IF{line-search}
        %     \STATE Compute \alpha_t^{(k)}\alpha_t^{(k)}
        % \ELSE
        %     \STATE \alpha_t^{(k)} = 1\alpha_t^{(k)} = 1
        % \ENDIF
        \STATE Update $\mathbf{\Delta u}_t^{(k+1)} =\mathbf{\Delta u}_t^{(k)}+\alpha_t^{(k)}\mathbf{p}_t^{(k)}$
        \STATE Update $\mathbf{u}_t^{(k+1)}=\mathbf{u}_t^{(k)}+\mathbf{\Delta u}_t^{(k+1)}$
    \ENDFOR
    \STATE $\mathbf{\Delta u}_t=\sum_{k=1}^K\mathbf{\Delta u}_t^{k+1}$
    \STATE $\mathbf{u}_t=\mathbf{u}_t^{(K+1)}\equiv \uref+\mathbf{\Delta u}_t$
\end{algorithmic}
\end{algorithm}
\section{Petrov-Galerkin projection}
We begin by approximating the solution state variable as a linear combination of $n$ orthogonal basis vectors $\boldsymbol{\phi}_i \in \mathbb{R}^N$ and coefficients $\deltauhat\in \mathbb{R}^{n}$. Here, variables denoted with a hat represent those of the reduced order dimension $n$. Consequently, we obtain

\begin{equation}
\deltatildeu_t(\boldsymbol{\mu}) = \mathbf{\Phi} \deltauhat_t(\boldsymbol{\mu}).
\end{equation}

This approximation enables us to update the solution, now referred to as $\tildeu_{t}$. The tilde notation indicates the approximate nature of this variable. Thus, the updated solution is given by

\begin{equation}
\tildeu_{t}= \uref+\mathbf{\Phi} \deltauhat_t(\boldsymbol{\mu}).
\label{Updated solution ROM}
\end{equation}
The right ROB $\mathbf{\Phi}$ is selected through Proper Orthogonal Decomposition (POD), which involves compressing a set of snapshots collected from the Full Order Model (FOM) simulations for carefully chosen parameters. Specifically, the finite element equations are solved for appropriately chosen values of the input parameter space $\mathcal{D}$, and the state variables are stored in snapshot matrices
\begin{equation}\mathbf{A}^u =
\begin{bmatrix}
\mathbf{u}_1(\boldsymbol{\mu}_1), &\mathbf{u}_2(\boldsymbol{\mu}_1),&\hdots,&\mathbf{u}_T(\boldsymbol{\mu}_1),&\mathbf{u}_1(\boldsymbol{\mu}_2), &\mathbf{u}_2(\boldsymbol{\mu}_2),&\hdots,& \mathbf{u}_T(\boldsymbol{\mu}_2),& \hdots,&\mathbf{u}_T(\boldsymbol{\mu}_P)
\end{bmatrix}\footnote{Although using $\mathbf{u}$ and $\mathbf{\Delta u}$ in snapshot generation theoretically results in the same column space under exact arithmetic or machine tolerance in randomized singular value decomposition (i.e., $col(\mathbf{A}^u)\subseteq col(\mathbf{A}^{\Delta u})$), we have presented the total approach in this work for its didactic value.}.
\end{equation}
The solution basis matrix $\mathbf{\Phi}$ is obtained as a linear combination of the columns of $\mathbf{A}^u$, i.e.,
\begin{equation}
    col(\mathbf{\Phi}) \subset col(\mathbf{A}^u)
\end{equation}
(the column space of $\mathbf{\Phi}$ is a subspace of the column space of $\mathbf{A}^u$). The Proper Orthogonal Decomposition (POD) method aims to reduce the number of columns in $\mathbf{\Phi}$ while preserving the essential patterns of the solution state variable. To achieve this, an error threshold $0 \leq \epsilon_u < 1$ is defined, and a basis matrix $\mathbf{\Phi}$ is sought such that:
\begin{equation}
\left\|\mathbf{A}^u-\mathbf{\Phi}\mathbf{\Phi}^T\mathbf{A}^u\right\|_F \leq \epsilon_u \left\|\mathbf{A}^u\right\|_F,
\end{equation}
where $\left\|\cdot\right\|_F$ denotes the Frobenius norm. One approach to finding this matrix is through truncated Singular Value Decomposition (SVD) \cite{Eckart1936} (see appendix \ref{app:Singular Value Decomposition}), which decomposes $\mathbf{A}^u$ as follows:
\begin{equation}
\mathbf{A}^u = \mathbf{\Phi}_n\mathbf{\Sigma}_n \mathbf{V}^T_n + \mathbf{E},
\end{equation}
where $\mathbf{\Phi}_n$ is a matrix containing the first $n$ left singular vectors of $\mathbf{A}^u$, $\mathbf{\Sigma}_n$ is a diagonal matrix containing the first $n$ singular values, and $\mathbf{V}_n$ is the matrix of the first $n$ right singular vectors. The truncation to $n$ modes allows for a low-rank approximation of $\mathbf{A}^u$, and the remaining term $\mathbf{E}$ accounts for the approximation error.\\ \\
When Eq.\ref{Updated solution ROM} is substituted into Eq.\ref{residual_FOM}, an over-determined system of $N$ nonlinear equations with $n$ unknowns is obtained, where $n<<N$, given by:
\begin{equation}
    \mathbf{R}_t(\tildeu_{t};\boldsymbol{\mu})=\mathbf{R}(\uref+\mathbf{\Phi \Deltaq}_t(\boldsymbol{\mu}),t;\boldsymbol{\mu})=\mathbf{0}.
    \label{Over-determined ROM}
\end{equation}
To obtain an approximate solution to this system of $N$ nonlinear equations with $n$ unknowns, $m$ constraints are imposed by requiring the orthogonality of the nonlinear residual to a left ROB $\mathbf{\Psi}\in\mathbb{R}^{N\times m}$. This process reduces the number of equations in the system from $N$ to $m$, resulting in a simplified system of $m$ state equations with $n$ unknowns given by
\begin{equation}
\mathbf{\Psi}_t^T\mathbf{R}_t(\uref+\mathbf{\Phi \Deltaq}_t(\boldsymbol{\mu});\boldsymbol{\mu})=\mathbf{0}.
    \label{Petrov-Galerkin system}
\end{equation}
Typically, the order of $m$ is similar to the order of $n$ ($\mathcal{O}(m)=\mathcal{O}(n)$), where $m<<N$. However, to avoid an under-determined system, it is necessary to ensure that $m$ is greater than or equal to $n$ ($m\geq n$) when reducing an over-determined system to a simplified system of $m$ state equations with $n$ unknowns, in order to ensure a well-posed problem. When $m\neq n$, the minimization problem can be addressed using QR decomposition. 
\\ \\
Applying Newton-Raphson to solve the reduced non-linear system to find a solution $\mathbf{u}_t$ results in the following iterations: for $k=1,\hdots,K$, solve
\begin{equation}
\boxed{\begin{aligned}
    \mathbf{\Psi}_t^{T(k)}\mathbf{J}_{t}^{(k)}\tildeutkm  \mathbf{\Phi} \hatptk & = -\mathbf{\Psi}_t^{T(k)}\mathbf{R}_t^{(k)}\tildeutkm\\
    \mathbf{\Deltaq}_t^{(k+1)} & =\mathbf{\Deltaq}_t^{(k)}+\alpha_t^{(k)}\hatptk\\
    \tildeu_t^{(k+1)}&=\tildeutk+\mathbf{\Phi}\mathbf{\Deltaq}_t^{(k+1)},
    \label{Newton-Raphson PG}
\end{aligned}}
\end{equation}
where $\hatptk \in \mathbb{R}^n$. The detailed procedure is given in Algorithm \ref{alg:Newton PG}.\\ \\
The quality of the approximate solution obtained is contingent on both the selection of the basis $\mathbf{\Psi}_t$ and the number of constraints imposed. A suitable basis, along with the optimal number of constraints, may be chosen to obtain a sufficiently accurate approximation of the original over-determined system of equations. Although other techniques based on Petrov-Galerkin frameworks have been introduced in the literature where the left ROB changes over iterations, i.e., $\mathbf{\Psi}_t^{(k)}$, this work presents a novel approach in section \ref{Invariant left ROB approach} for a Petrov-Galerkin framework that finds an invariant or constant left ROB, i.e. $\mathbf{\Psi}$, while preserving the minimum-residual optimality that will be discussed in the following section.
\begin{algorithm}
\caption{Newton-Raphson strategy for general PG-PROM.}
\label{alg:Newton PG}
\hspace*{\algorithmicindent} \textbf{Input:} State variable $\uref$ and right  ROB $\mathbf{\Phi}$\\
\hspace*{\algorithmicindent} \textbf{Output:} State variable $\tildeu_t$ and increment update $\deltatildeu_t$
\begin{algorithmic}[1]
    \STATE Initialize $\mathbf{\Deltaq}_t^{(0)}=\mathbf{0}$, i.e., $\tildeu_t^{(0)}=\uref$
    \FOR{$k=0,\hdots,K $ (convergence)}
        \STATE Evaluate residual $\mathbf{R}_t^{(k)}=\left.\mathbf{R}_t\right|_{\tildeutk}$
        \STATE Evaluate Jacobian $\mathbf{J}_t^{(k)}=\left.\frac{\partial \mathbf{R}_t}{\partial \mathbf{\tildeu}}\right|_{\tildeutk}$
        \STATE Compute $\mathbf{\Psi}_t^{T(k)}$ (see section \ref{Optimal left ROB})
        \STATE Compute $\mathbf{W}^{(k)} = \mathbf{\Psi}_t^{T(k)}\mathbf{J}_{t}^{(k)}\mathbf{\Phi}$
        \IF{$m\neq n$}
            \STATE Compute QR decomposition $\mathbf{W}^{(k)} = \mathbf{Q}^{(k)} \mathbf{D}^{(k)}$
            \STATE Solve $\mathbf{D}^{(k)}  \hatptk = -\mathbf{Q}^{T(k)}\mathbf{\Psi}_t^{T(k)}\mathbf{R}_t^{(k)}$
        \ELSE
            \STATE Solve $\mathbf{W}^{(k)}  \hatptk = -\mathbf{\Psi}_t^{T(k)}\mathbf{R}_t^{(k)}$
        \ENDIF
        % \IF{line-search}
        %     \STATE Compute \alpha_t^{(k)}\alpha_t^{(k)}
        % \ELSE
        %     \STATE \alpha_t^{(k)} = 1\alpha_t^{(k)} = 1
        % \ENDIF
        \STATE Update $\mathbf{\Deltaq}_t^{(k+1)} =\mathbf{\Deltaq}_t^{(k)}+\alpha_t^{(k)}\hatptk$
        \STATE Update $\tildeu_t^{(k+1)}=\tildeutk+\mathbf{\Phi}\mathbf{\Deltaq}_t^{(k+1)}$
    \ENDFOR
    \STATE $\deltatildeu_t=\mathbf{\Phi}\sum_{k=1}^K\mathbf{\Deltaq}_t^{k+1}$
    \STATE $\tildeu_t=\tildeu_t^{(K+1)}\equiv \uref+\deltatildeu_t$
\end{algorithmic}
\end{algorithm}
\subsection{Optimal left ROB}
\label{Optimal left ROB}
The left ROB $\mathbf{\Psi}_t$ is chosen here to minimize the non-linear residual in some norm $\mathbf{G}_t$ in order to meet the minimum-residual optimality of the projection approximation. Specifically, we seek to solve the following minimization problem:
\begin{equation}
    \min_{\mathbf{\Deltaq}_t \in \mathbb{R}^n} \frac{1}{2}\norm\Big{\mathbf{R}_t(\uref+\mathbf{\Phi}  \mathbf{\Deltaq}_t(\boldsymbol{\mu});\boldsymbol{\mu})}_{\mathbf{G}_t}^2 
    \label{Minimization of the residual}
\end{equation}
where $\|\cdot\|_{\mathbf{G}_t}$ denotes a norm defined by a symmetric, positive definite (SPD) matrix $\mathbf{G}_t \in \mathbb{R}^{N\times N}$. Alternatively, we can write the above problem as follows:
\begin{equation}
    \min_{\mathbf{\Deltaq}_t \in \mathbb{R}^n} \frac{1}{2}\mathbf{R}_t\tildeutm^T\mathbf{G}_t\mathbf{R}_t\tildeutm.
\end{equation}
By satisfying the minimum-residual optimality of Eq.\ref{Minimization of the residual}, a general relationship between projection-based reduced-order models and minimum-residual reduced-order models can be established, as follows: 
\begin{equation}
\begin{aligned}
    \left[\frac{\partial \mathbf{R}_t\tildeutm}{\partial \tildeu} \frac{\partial \tildeu}{\partial \mathbf{\Deltaq}_t}\right]^T\mathbf{G}_t\mathbf{R}_t\tildeutm&=\mathbf{0}\\
    \left[\mathbf{J}_t\tildeutm\mathbf{\Phi}\right]^T\mathbf{G}_t\mathbf{R}_t\tildeutm&=\mathbf{0},
\end{aligned}
\end{equation}
with the structure of the PROM in Eq.\ref{Petrov-Galerkin system}; this yields
\begin{equation}
    \mathbf{\Psi}_t = \mathbf{G}_t\mathbf{J}_t\tildeutm\mathbf{\Phi},
    \label{General left ROB}
\end{equation}
which is sufficient to satisfy the minimum-residual property for a general PROM. For further details, please refer to \cite{zahr2016phd}.
\subsubsection{Galerkin projection}
The Galerkin projection produces search directions $\hatptk$ that are minimum-residual optimal with $\mathbf{G}_t=\mathbf{J}_t^{-1}$ when the Jacobians are SPD matrices; that is,
\begin{equation}
    \mathbf{\Psi}_t\tildeutm = \mathbf{J}_t^{-1}\tildeutm\mathbf{J}_t\tildeutm\mathbf{\Phi}= \mathbf{\Phi},
\end{equation}
where the right ROB $\mathbf{\Phi}$ is constant. Consequently, the left ROB no longer depends on time and the parameter, becoming $\mathbf{\Psi}_t\tildeutm =\mathbf{\Psi}$, see Ref.\cite{Saad2003}.
The Newton-Raphson system is then transformed into the following iterations: for $k=1,\hdots,K$, solve
\begin{equation}
\boxed{\begin{aligned}
    \mathbf{\Phi}^{T}\mathbf{J}_{t}^{(k)}\tildeutkm \mathbf{\Phi}  \hatptk & = -\mathbf{\Phi}^{T}\mathbf{R}_t^{(k)}\tildeutkm\\
    \mathbf{\Deltaq}_t^{(k+1)} & =\mathbf{\Deltaq}_t^{(k)}+\alpha_t^{(k)}\hatptk\\
    \tildeu_t^{(k+1)}&=\tildeutk+\mathbf{\Phi}\mathbf{\Deltaq}_t^{(k+1)},
    \label{Newton-Raphson Galerkin}
\end{aligned}}
\end{equation}
see algorithm \ref{alg:Newton Galerkin} for more details.
\\ 
However, the Galerkin projection does not guarantee that the residual is minimized since the Jacobians of a nonlinear problem are not in general SPD matrices. As a result, a different projection for general non-linear problems was introduced in Ref.\cite{Carlberg2011}, and is discussed in the following section.
\begin{algorithm}
\caption{Newton-Raphson strategy for Galerkin-PROM.}
\label{alg:Newton Galerkin}
\hspace*{\algorithmicindent} \textbf{Input:} State variable $\uref$ and right  ROB $\mathbf{\Phi}$\\
\hspace*{\algorithmicindent} \textbf{Output:} State variable $\tildeu_t$ and increment update $\deltatildeu_t$
\begin{algorithmic}[1]
    \STATE Initialize $\mathbf{\Deltaq}_t^{(0)}=\mathbf{0}$, i.e., $\tildeu_t^{(0)}=\uref$
    \FOR{$k=0,\hdots,K $ (convergence)}
        \STATE Evaluate residual $\mathbf{R}_t^{(k)}=\left.\mathbf{R}_t\right|_{\tildeutk}$
        \STATE Evaluate Jacobian $\mathbf{J}_t^{(k)}=\left.\frac{\partial \mathbf{R}_t}{\partial \mathbf{\tildeu}}\right|_{\tildeutk}$
        \STATE Compute $\mathbf{W}^{(k)} = \mathbf{\Phi}^{T}\mathbf{J}_{t}^{(k)}\mathbf{\Phi}$
        \STATE Solve $\mathbf{W}^{(k)}  \hatptk = -\mathbf{\Phi}^{T}\mathbf{R}_t^{(k)}$
        % \IF{line-search}
        %     \STATE Compute \alpha_t^{(k)}\alpha_t^{(k)}
        % \ELSE
        %     \STATE \alpha_t^{(k)} = 1\alpha_t^{(k)} = 1
        % \ENDIF
        \STATE Update $\mathbf{\Deltaq}_t^{(k+1)} =\mathbf{\Deltaq}_t^{(k)}+\alpha_t^{(k)}\hatptk$
        \STATE Update $\tildeu_t^{(k+1)}=\tildeutk+\mathbf{\Phi}\mathbf{\Deltaq}_t^{(k+1)}$
    \ENDFOR
    \STATE $\deltatildeu_t=\mathbf{\Phi}\sum_{k=1}^K\mathbf{\Deltaq}_t^{k+1}$
    \STATE $\tildeu_t=\tildeu_t^{(K+1)}\equiv \uref+\deltatildeu_t$
\end{algorithmic}
\end{algorithm}
\subsubsection{Least-Squares Petrov-Galerkin (LSPG) projection}
Nonlinear problems often result in Jacobians that are not SPD. In such cases, the LSPG method satisfies the condition in Eq.\ref{General left ROB} by setting $\mathbf{G}_t = \mathbf{I}$, where $\mathbf{I} \in \mathbb{R}^{N\times N}$ is the identity matrix. This ensures that the method exhibits the minimum-residual property, leading to the following equation:

\begin{equation}
\mathbf{\Psi}_t \tildeutm= \mathbf{J}_t\tildeutm\mathbf{\Phi}.
\end{equation}
The above equation leads to the Gauss-Newton iterations, which are given by the following steps: for $k = 1,\dots,K$, solve
\begin{equation}
\boxed{\begin{aligned}
    \underbrace{\left[\mathbf{J}_{t}^{(k)}\tildeutkm \mathbf{\Phi}\right]^T}_{\mathbf{\Psi}_t^{T(k)}\tildeutkm}\mathbf{J}_{t}^{(k)}\tildeutkm \mathbf{\Phi}  \hatptk & = -\underbrace{\left[\mathbf{J}_{t}^{(k)}\tildeutkm \mathbf{\Phi}\right]^T}_{\mathbf{\Psi}_t^{T(k)}\tildeutkm}\mathbf{R}_t^{(k)}\tildeutkm\\
    \mathbf{\Deltaq}_t^{(k+1)} & =\mathbf{\Deltaq}_t^{(k)}+\alpha_t^{(k)}\hatptk\\
    \tildeu_t^{(k+1)}&=\tildeutk+\mathbf{\Phi}\mathbf{\Deltaq}_t^{(k+1)}.
    \label{Newton-Raphson LSPG}
\end{aligned}}
\end{equation}
Note that the chosen left ROB $\mathbf{\Psi}_t \tildeutm= \mathbf{J}_t\tildeutm\mathbf{\Phi}$ changes from one Newton iteration to another (see algorithm \ref{alg:Newton LSPG} for more details).\\ \\
\begin{algorithm}
\caption{Gauss-Newton strategy for LSPG-PROM.}
\label{alg:Newton LSPG}
\hspace*{\algorithmicindent} \textbf{Input:} State variable $\uref$ and right  ROB $\mathbf{\Phi}$\\
\hspace*{\algorithmicindent} \textbf{Output:} State variable $\tildeu_t$ and increment update $\deltatildeu_t$
\begin{algorithmic}[1]
    \STATE Initialize $\mathbf{\Deltaq}_t^{(0)}=\mathbf{0}$, i.e., $\tildeu_t^{(0)}=\uref$
    \FOR{$k=0,\hdots,K $ (convergence)}
        \STATE Evaluate residual $\mathbf{R}_t^{(k)}=\left.\mathbf{R}_t\right|_{\tildeutk}$
        \STATE Evaluate Jacobian $\mathbf{J}_t^{(k)}=\left.\frac{\partial \mathbf{R}_t}{\partial \mathbf{\tildeu}}\right|_{\tildeutk}$
        \STATE Compute left ROB $\mathbf{\Psi}_t^{T(k)}=\left[\mathbf{J}_t^{(k)}\mathbf{\Phi}\right]^T$
        \STATE Compute $\mathbf{W}^{(k)} = \mathbf{\Psi}_t^{T(k)}\mathbf{J}_{t}^{(k)}\mathbf{\Phi}$
        \STATE Solve $\mathbf{W}^{(k)}  \hatptk = -\mathbf{\Psi}_t^{T(k)}\mathbf{R}_t^{(k)}$
        % \IF{line-search}
        %     \STATE Compute \alpha_t^{(k)}\alpha_t^{(k)}
        % \ELSE
        %     \STATE \alpha_t^{(k)} = 1\alpha_t^{(k)} = 1
        % \ENDIF
        \STATE Update $\mathbf{\Deltaq}_t^{(k+1)} =\mathbf{\Deltaq}_t^{(k)}+\alpha_t^{(k)}\hatptk$
        \STATE Update $\tildeu_t^{(k+1)}=\tildeutk+\mathbf{\Phi}\mathbf{\Deltaq}_t^{(k+1)}$
    \ENDFOR
    \STATE $\deltatildeu_t=\mathbf{\Phi}\sum_{k=1}^K\mathbf{\Deltaq}_t^{k+1}$
    \STATE $\tildeu_t=\tildeu_t^{(K+1)}\equiv \uref+\deltatildeu_t$
\end{algorithmic}
\end{algorithm}
Nonlinear PROMs typically involve at least one step that increases in computational complexity with the size of the FOM, even though the nonlinear system that defines the reduced-order model is much smaller. To overcome this bottleneck, many nonlinear PROM methods include a hyper-reduction technique that adds an additional approximation. For instance, in Newton-Raphson methods, the residual vector $\mathbf{R}_t^{(k)}\tildeutm$, Jacobian matrix $\mathbf{J}_t^{(k)}\tildeutm$, and matrix-vector products for all entities or elements of the FOM are evaluated at each iteration for all time steps. Hyper-reduction techniques can tackle this bottleneck by evaluating these vectors and matrices over a subset of elements, significantly reducing the computational burden without compromising accuracy. It is worth noting that if the left ROB is variant or iteration-dependent ($\mathbf{\Psi}_t^{(k)}\tildeutm$), it may pose a disadvantage to the ``hyper-reduction" strategy. In such cases, constructing the PROM element by element becomes infeasible, and a complementary mesh needs to be used for the reduced mesh. Details regarding the need for a complementary mesh approach are discussed in the following section.
\section{Hyper-reduction}
\label{Hyper-reduction}
To discuss hyper-reduction, let us first revisit Eq. \ref{Petrov-Galerkin system} for convenience:
\begin{equation*}
\mathbf{\Psi}_t^T\mathbf{R}_t(\uref+\mathbf{\Phi \Deltaq}_t(\boldsymbol{\mu});\boldsymbol{\mu})=\mathbf{0}.
\end{equation*}
For the assembly of elemental contributions, this expression takes the following form:
\begin{equation}
    \mathbf{\Psi}_t^T\mathbf{R}_t=\mathbf{\Psi}_t^T\sum_{e=1}^{L}\mathbf{L}^{e^T}\mathbf{R}_t^{e}=\sum_{e=1}^{L}(\mathbf{L}^{e}\mathbf{\Psi}_t)^T\mathbf{R}_t^{e}=\sum_{e=1}^{L}\mathbf{\Psi}_t^{e^T}\mathbf{R}_t^{e}=\mathbf{0},
    \label{Galerkin assembly of elemental contributions}
\end{equation}
where $L$ denotes the total number of elements, and for each element, $\mathbf{L}^e$ is the assembly operator, $\mathbf{R}_t^e$ is the elemental residual vector, and $\mathbf{\Psi}_t^e$ represents the values of the left ROB associated with the elemental degrees of freedom. As previously mentioned, this equation shows that although the equation-solving effort has drastically diminished with $n << N$, the computational complexity of the problem still scales with the size of the discretization of the underlying finite elements. This is because it is necessary to loop over the $L$ finite elements of the mesh to assemble the residual vectors and Jacobian matrices.\\ \\
As demonstrated in \cite{Joaquin2020, Farhat2015}, one may approximate the Petrov-Galerkin projection of the residual onto the left ROB, given by Eq.\ref{Galerkin assembly of elemental contributions}, by finding a subset of elements $\mathbf{z}$ and positive weights $\boldsymbol{\omega}$ such that
\begin{equation}
    \sum_{e\in \mathbf{z}}\omega^e\mathbf{\Psi}_t^{e^T}\mathbf{R}_t^e=\mathbf{0}.
\end{equation}
In other words, the subset of elements and their weights are chosen such that the resulting approximation of the residual vector satisfies the projection property of the reduced order model. Specifically, the approximation of the residual vector is obtained by summing over the chosen elements in $\mathbf{z}$, where for each element $e$ in $\mathbf{z}$, the corresponding weight $\omega^e \in \boldsymbol{\omega}$ is multiplied by the residual vector $\mathbf{R}_t^e$ and its corresponding left ROB values $\mathbf{\Psi}_t^{e^T}$. 
\\
\\
The optimization problem for determining $\mathbf{z}$ and $\boldsymbol{\omega}$ involves solving Eq.\ref{Galerkin assembly of elemental contributions} using the same input parameters as in the first reduction stage. The reduced equation is expressed in matrix format as follows:

\begin{equation}
    \sum_{e=1}^L\mathbf{\Psi}_i^{e^T}\mathbf{R}_i^e(\boldsymbol{\mu}_j)=
    \begin{bmatrix}
    \mathbf{\Psi}^{1^T}\mathbf{R}_i^1(\boldsymbol{\mu}_j), & \mathbf{\Psi}^{2^T}\mathbf{R}_i^2(\boldsymbol{\mu}_j), & \hdots, & \mathbf{\Psi}^{L^T}\mathbf{R}_i^L(\boldsymbol{\mu}_j)
    \end{bmatrix}\mathbbm{1}=\mathbf{b}_i(\boldsymbol{\mu}_j),
    \qquad \qquad \qquad
    i=1,2,\hdots,T,
    \label{Projected Residuals}
\end{equation}
where $\mathbf{b}_i(\boldsymbol{\mu}_j)\in\mathbb{R}^m$ represents the sum of the projected residuals for all elements at the $i^{th}$ time step for the $j^{th}$ parameter instance. Here, $\mathbbm{1}$ denotes an all-ones vector. 
\begin{remark}
Using the all-ones vector assumes that the mesh elements are similar enough to each other, such that each element's contribution to the projected residual is approximately equal. An alternative approach is to create a weighting vector that incorporates a coefficient to account for the varying contributions, such as the volume of each element \cite{Joaquin2016}. If a weighting vector is used, then the residual should be scaled component-wise by its corresponding weight.
\end{remark}
Let us construct a matrix $\mathbf{X}\in \mathbb{R}^{(m\times T\times P)\times L}$ by collecting the block column matrix from Eq.\ref{Projected Residuals} for all training parameters, which is given by:
\begin{equation}
    \mathbf{X}=
    \begin{bmatrix}
    \mathbf{X}_1(\boldsymbol{\mu}_1),&\mathbf{X}_2(\boldsymbol{\mu}_1),&\hdots&\mathbf{X}_T(\boldsymbol{\mu}_1),&\mathbf{X}_1(\boldsymbol{\mu}_2),&\mathbf{X}_2(\boldsymbol{\mu}_2),&\hdots,&\mathbf{X}_T(\boldsymbol{\mu}_2),&\hdots,&\mathbf{X}_T(\boldsymbol{\mu}_P)
    \end{bmatrix}^T,
\end{equation}
where each block column $\mathbf{X}_i(\boldsymbol{\mu}_j)\in \mathbb{R}^{L\times m}$ is given by:
\begin{equation}
    \mathbf{X}_i(\boldsymbol{\mu}_j)=
    \begin{bmatrix}
    (\mathbf{\Psi}^{1^T}\mathbf{R}_i^1(\boldsymbol{\mu}_j))^T\\
    (\mathbf{\Psi}^{2^T}\mathbf{R}_i^2(\boldsymbol{\mu}_j))^T\\
    \vdots\\
    (\mathbf{\Psi}^{L^T}\mathbf{R}_i^L(\boldsymbol{\mu}_j))^T\\
    \end{bmatrix}.
\end{equation}
The discrete minimization problem statement is very similar to the statement for the continuous problem in Ref.\cite{An2008}. The objective is to determine the optimal subset of elements and their corresponding weights. Therefore, we need to find $\boldsymbol{\omega}\in\mathbb{R}^{\ell}_+$ and $\mathbf{z}\subset {1,2,\hdots,L}$ such that:
\begin{equation}
    (\boldsymbol{\omega},\mathbf{z})= \arg \min_{\boldsymbol{\bar{\omega}\geq \mathbf{0},\mathbf{\bar{z}}}}\|\mathbf{X_{\bar{z}}}\boldsymbol{\bar{\omega}}-\mathbf{b}\|^2,
    \label{ECM minimization}
\end{equation}
where $\mathbf{X}_{\mathbf{z}}$ is the block matrix of $\mathbf{X}$ formed by the columns corresponding to the indexes $\mathbf{z}$, and $\boldsymbol{\omega}=[\omega^1,\omega^2,\hdots,\omega^{\ell}]^T$. The goal is to select a set of $\ell$ elements from the $L$ elements of the underlying finite element mesh ${x_1, x_2,..., x_L}$. For more information on the ``Empirical Cubature Method" (ECM) for finding the optimal elements and weights, we refer the reader to Refs.\cite{Joaquin2016,Joaquin2020}.
Unlike other methods in the literature \cite{Farhat2015, Grimberg2021}, the Empirical Cubature Method solves the minimization problem in Eq.\ref{ECM minimization} in terms of a set of orthogonal basis vectors rather than snapshots of the raw discrete integrand $\mathbf{X}$. To this end, the ECM requires obtaining an orthonormal matrix via a truncated SVD as
\begin{equation}
    \mathbf{X}=\mathbf{U}\boldsymbol{\Sigma}\boldsymbol{\Theta}+\mathbf{E}.
\end{equation}
The matrix $\boldsymbol{\Theta}\in \mathbb{R}^{p\times L}$ is used to define a vector $\mathbf{b}_{\boldsymbol{\Theta}}\in \mathbb{R}^L$ as
\begin{equation}
    \boldsymbol{\Theta}\mathbbm{1}=\mathbf{b}_{\boldsymbol{\Theta}},
\end{equation}
such that the equivalent optimization problem statement in Eq.\ref{ECM minimization} becomes:
\begin{equation}
    (\boldsymbol{\omega},\mathbf{z})= \arg \min_{\boldsymbol{\bar{\omega}\geq \mathbf{0},\mathbf{\bar{z}}}}\|\boldsymbol{\Theta}_{\mathbf{\bar{z}}}\boldsymbol{\bar{\omega}}-b_{\boldsymbol{\Theta}}\|^2.
\end{equation}
The ECM algorithm for finding the optimal subset of elements and weights is described in detail in \cite{Joaquin2020}, and is used in this work to address the hyper-reduction strategy. Note that the hyper-reduction approximation does not aim to accurately represent the full-order residual (FOM), but only its projection onto the left subspace spanned by the columns of the left reduced order basis (ROM). \\
\\
The hyper-reduction technique has been extensively studied in previous literature for the Galerkin projection \cite{Chaturantabut2010, Farhat2015, Galbally2010, Joaquin2016, Joaquin2020, Ryckelynck2005, Barrault2004, Farhat2014}. However, since it is a relatively new approach, only a limited number of studies have investigated scenarios where the left reduced-order basis (ROB) differs \cite{Carlberg2011, Grimberg2021, Fang2013, Shimizu2021}. This introduces some unique challenges to the hyper-reduction scheme.

\subsection{Extension to least-squares problems}
\label{Extension to least-squares problems}
Let us consider the case when the minimization is performed in the 2-norm, which corresponds to $\mathbf{G}_t=\mathbf{I}$. In this case, Eq.\ref{Minimization of the residual} becomes
\begin{equation}
    \mathbf{\Phi}^T\mathbf{J}_t^T\tildeutm\mathbf{R}_t\tildeutm=\mathbf{0}.
    \label{Least Squares projected residual}
\end{equation}
We define the Jacobian-weighted residual vector as follows:
\begin{equation}
    \Bar{\mathbf{R}}_t\tildeutm=\mathbf{J}_t^T\mathbf{R}_t\tildeutm=\mathbf{0}.
\end{equation}
With this definition at hand, we can rewrite Eq.\ref{Least Squares projected residual} as
\begin{equation}
    \mathbf{\Phi}^T \Bar{\mathbf{R}}_t\tildeutm=\mathbf{0},
    \label{Least Squares projected residual modified}
\end{equation}
which, except for the $\Bar{\mathbf{R}}_t$ operator, is identical to Eq.\ref{Petrov-Galerkin system} in the sense that it involves a ROB-weighted residual quantity. However, because the Jacobian-weighted residual vector involves global quantities, it cannot be assembled locally element by element. This means that it requires information from a ``complementary mesh", which is an additional set of elements used to accurately represent the global behavior of the system on the reduced mesh.

\begin{figure}[H]
\centering
\fbox{\includegraphics[width=0.4\linewidth]{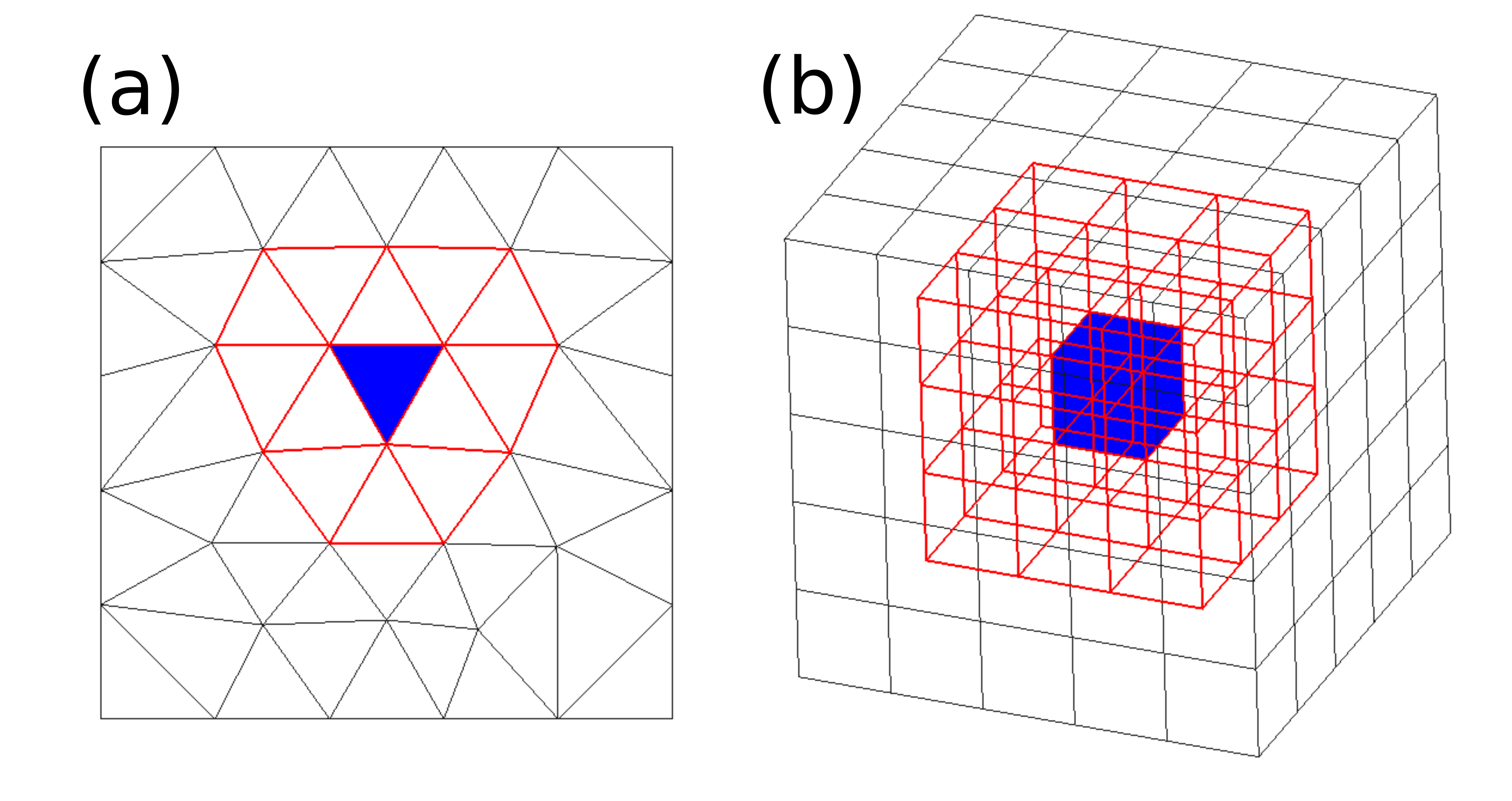}}
\caption{Patch of elements associated to a selected element of a 2D and 3D mesh. (a) 2D, (b) 3D. Red: Patch of elements, Blue: Selected element}
\label{fig:Patch of elements associated to a selected element of a 2D and 3D mesh.}
\end{figure}

To further clarify this point, we can express Eq.\ref{Least Squares projected residual modified} in terms of elemental contributions as follows:
\begin{equation}
\begin{split}
    \mathbf{\Phi} ^T\mathbf{J}_t^T\mathbf{R}_t&=\mathbf{\Phi}^T\left(\sum_{e=1}^L\mathbf{L}^{e^T}\mathbf{J}_t^{e^T}\mathbf{L}^{e}\right)\mathbf{R}_t\\
    & = (\sum_{e=1}^L\left(\mathbf{L}^{e}\mathbf{\Phi}\right)^T\mathbf{J}_t^{e^T}\mathbf{L}^{e})\mathbf{R}_t\\
    & = \left(\sum_{e=1}^L\mathbf{\Phi}^{e^T}\mathbf{J}_t^{e^T}\mathbf{L}^{e}\right)\mathbf{R}_t,\\
    \label{Leasts Squares Projected Residuals}
\end{split}
\end{equation}
where we introduce the elemental quantity
\begin{equation}
\rtle := \mathbf{L}^e \mathbf{R}_t.
\label{Assembled residual vector}
\end{equation}
The quantity $\rtle$ signifies the entries of the assembled residual vector that correspond to the degrees of freedom of element $e$. It is important not to confuse this quantity with $\mathbf{R}^e$, which denotes the contribution of element $e$ to the residual vector. To emphasize this distinction, note that $\rtle$ and $\mathbf{R}^e$ are not the same, i.e., $\rtle\neq \mathbf{R}^e$.\\
Therefore, we have
\begin{equation}
    \sum_{e=1}^L\mathbf{\Phi}^{e^T}\barrte\tildeutm=\mathbf{0},
\end{equation}
where
\begin{equation}
    \barrte:=\mathbf{J}_t^{e^T}\rtle.
    \label{Elemental Residual multiplied by the elemental jacobian}
\end{equation}
The main difference between the LSPG scheme and other schemes, such as Galerkin, is that the computation of $\barrte$ in the former requires determining residual vectors for all elements sharing nodes with element $e$, forming a patch of elements, as shown in Fig. \ref{fig:Patch of elements associated to a selected element of a 2D and 3D mesh.}. This feature necessitates constructing a complementary mesh based on the patch of elements for a given set of elements $\mathbf{z}\subset {1,2,\hdots,L}$, regardless of the element selection method employed in the hyper-reduction process \cite{Grimberg2021}. This peculiarity of problems arising from minimization of residuals suggests that the ECM search process should be adjusted such that, after selecting a specific element, its associated patches are considered as candidates for the next step in the algorithm. This maximizes the overlap between patches. Using the hyper-reduction scheme described in Section \ref{Hyper-reduction} for the variant left ROB would result in an increased number of elements to consider in the finite element analyses.
\section{Invariant left ROB approach}
\label{Invariant left ROB approach}
\subsection{Formulation}
Let us revisit Eq.\ref{Least Squares projected residual}, which is reproduced below for convenience (in the case where $\mathbf{G}_t = \mathbf{I}$):
\begin{equation*}
    \mathbf{r}_t = \mathbf{\Phi}^T\mathbf{J}_t^T\tildeutm\mathbf{R}_t\tildeutm = \mathbf{0}.
    \label{Derivative Objective Function LS}
\end{equation*}
This nonlinear equation can be interpreted as a Petrov-Galerkin projection in which the left subspace or the subspace of constraints (a term borrowed from \cite{Saad2003}) changes at each iteration. The left subspace in the preceding equation is given by $col(\mathbf{J}_t\mathbf{\Phi})$, where $\mathbf{J}_t$ is the Jacobian of the original finite element equations, and this matrix evolves during the iterations. In light of the preceding observation, one may wonder whether it is possible to reformulate Eq.\ref{Least Squares projected residual} so that the left ROB matrix is held fixed during the iterations.\\
\\
The problem may be posed as follows: find an orthogonal matrix $\mathbf{\Psi} \in \mathbb{R}^{N\times m}$ such that
\begin{equation}
    \mathbf{\Psi}^T\mathbf{R}_t\tildeutm=\mathbf{0}
    \label{Petrov Galerkin constrained Residual}
\end{equation}
for all input parameters $\boldsymbol{\mu}$ used for determining the right ROB $\mathbf{\Phi} \in \mathbb{R}^{N\times n}$.\\
\\
In what follows, we describe and discuss two possible approaches for determining the left ROB $\mathbf{\Psi}$. These two alternatives will introduce a second training stage in which data will be gathered from the LSPG-PROM solution, leading to an increase in offline computational cost. However, we demonstrate in the sequel that this increase in computational effort is greatly repaid in the online stage, as the resulting PG-PROM requires far fewer elements than the standard LSPG-PROM.
\subsubsection{Jacobian-based approach}
\label{Jacobian-based approach}
\begin{framed}
\textit{Jacobian basis procedure}\\
\begin{enumerate}
    \item This approach involves solving Eq.\ref{Derivative Objective Function LS} while collecting the matrices $\mathbf{J}_i(\tildeu_i;\boldsymbol{\mu}_j)\mathbf{\Phi}$ when convergence is achieved (for all training parameters $\boldsymbol{\mu})$. 

First, define the auxiliary matrix:

\begin{equation}
\mathbf{S}_{\mathbf{J}}^j(\boldsymbol{\mu}_j) :=
\begin{bmatrix}
\mathbf{J}_1(\tildeu_1;\boldsymbol{\mu}_j)\mathbf{\Phi}, & \mathbf{J}_2(\tildeu_2;\boldsymbol{\mu}_j)\mathbf{\Phi}, & \hdots, & \mathbf{J}_T(\tildeu_T;\boldsymbol{\mu}_j)\mathbf{\Phi}
\end{bmatrix}.
\end{equation}

Then construct $\mathbf{S}_{\mathbf{J}}$ by stacking these matrices:

\begin{equation}
\mathbf{S}_{\mathbf{J}} =
\begin{bmatrix}
\mathbf{S}_{\mathbf{J}}^1(\boldsymbol{\mu}_1), &
\mathbf{S}_{\mathbf{J}}^2(\boldsymbol{\mu}_2), &
\hdots, &
\mathbf{S}_{\mathbf{J}}^P(\boldsymbol{\mu}_P)
\end{bmatrix}.
\label{Snapshots Matrix Jacobian}
\end{equation}
\item  Once the matrix has been constructed, the truncated SVD (with relative truncation tolerance $\epsilon_{\mathbf{\Psi}_{\mathbf{J}}})$ is used to determine an orthogonal basis matrix: 
\begin{equation}
\left[\mathbf{\Psi}_{\mathbf{J}}, \bullet, \bullet\right] =\text{SVD}(\mathbf{S}_{\mathbf{J}},\epsilon_{\mathbf{\Psi}_{\mathbf{J}}}).
\label{Jacobian SVD}
\end{equation}
\end{enumerate}
\end{framed}
It is worth noting that although all iterative schemes in this work are presented using the Newton-Raphson method, the proposed approach is not restricted to this method and can be applied to any iterative scheme, including those that use fixed-point iteration techniques such as Picard's method. Therefore, the matrices $\mathbf{J}_i$ can be computed using either the exact Jacobian or an approximate Jacobian. \\
\textbf{Observations:}
\begin{enumerate}
    \item The number of columns of $\mathbf{\Psi}_{\mathbf{J}}$ is always equal to or greater than the number of columns of $\mathbf{\Phi}$:
    \begin{equation}
        ncol(\mathbf{\Psi}_{\mathbf{J}})\geq ncol(\mathbf{\Phi}).
    \end{equation}
    The equality holds in the linear case, i.e., when $\mathbf{J}_t$ is constant.
    \item As a corollary, in the general case  where $ncol(\mathbf{\Psi}_{\mathbf{J}}) > ncol(\mathbf{\Phi})$, Eq.\ref{Petrov Galerkin constrained Residual} becomes an over-determined system of nonlinear equations. This implies that the problem of Eq.\ref{Petrov Galerkin constrained Residual} should be posed as finding $\mathbf{\Deltaq}_t \in \mathbb{R}^n$ such that
    \begin{equation}
        \min_{\mathbf{\Deltaq}_t\in\mathbb{R}^n}\frac{1}{2} \norm\Big{\mathbf{\Psi}_{\mathbf{J}}^T \mathbf{R}_t\tildeutm}^2,
    \end{equation}
    which in turn, leads to the stationary condition
    
    \begin{equation}
    \begin{aligned}
        \frac{\partial{\frac{1}{2} (\mathbf{\Psi}_{\mathbf{J}}^T \mathbf{R}_t)^T(\mathbf{\Psi}_{\mathbf{J}}^T \mathbf{R}_t)}}{\partial{\mathbf{\Deltaq}_t}} &= \frac{1}{2}(\mathbf{\Psi}_{\mathbf{J}}^T \frac{\partial \mathbf{R}_t}{\partial \mathbf{\Deltaq}_t})^T(\mathbf{\Psi}_{\mathbf{J}}^T \mathbf{R}_t) + \frac{1}{2}(\mathbf{\Psi}_{\mathbf{J}}^T \mathbf{R}_t)^T(\mathbf{\Psi}_{\mathbf{J}}^T \frac{\partial \mathbf{R}_t}{\partial \mathbf{\Deltaq}_t}) \\
        &=\frac{1}{2}(\mathbf{\Psi}_{\mathbf{J}}^T\mathbf{J}_t\mathbf{\Phi})^T(\mathbf{\Psi}_{\mathbf{J}}^T \mathbf{R}_t) + \frac{1}{2}(\mathbf{\Psi}_{\mathbf{J}}^T \mathbf{R}_t)^T(\mathbf{\Psi}_{\mathbf{J}}^T\mathbf{J}_t\mathbf{\Phi}) \\
        &= (\mathbf{\Psi}_{\mathbf{J}}^T\mathbf{J}_t\mathbf{\Phi})^T(\mathbf{\Psi}_{\mathbf{J}}^T \mathbf{R}_t) = \mathbf{0}.
    \end{aligned}
    \end{equation}
    Hence, in this case, the residual adopts the expression
    \begin{equation}
        \mathbf{r}'_t=\mathbf{\Phi}^T(\mathbf{J}_t^T\mathbf{\Psi}_{\mathbf{J}})(\mathbf{\Psi}_{\mathbf{J}}^T\mathbf{R}_t)=\mathbf{0}.
        \label{Prime residual}
    \end{equation}
\end{enumerate}
If we use the standard Newton-Raphson method to solve the preceding equation, obtain the following system of linear equations at each iteration:
\begin{equation}
    \frac{\partial{\mathbf{r}'}}{\partial{\mathbf{\Deltaq}_t}} \hatpt=-\mathbf{r}'_t,
    \label{Newton-Rapshon prime residual}
\end{equation}
where the Jacobian matrix of the projected residual $\mathbf{r}'$ is given by

\begin{equation}
\frac{\partial{\mathbf{r}'_t}}{\partial{\mathbf{\Deltaq}_t}} = \mathbf{\Phi}^T\frac{\partial{\mathbf{J}_t^T}}{\partial{\mathbf{\Deltaq}_t}}\mathbf{\Psi}_{\mathbf{J}}\mathbf{\Psi}_{\mathbf{J}}^T \mathbf{R}_t + \mathbf{\Phi}^T\mathbf{J}_t^T\mathbf{\Psi}_{\mathbf{J}}\mathbf{\Psi}_{\mathbf{J}}^T \frac{\partial{\mathbf{R}_t}}{\partial{\mathbf{\Deltaq}_t}}
\end{equation}
Near the minimum, i.e., when $\mathbf{R}_t \approx \mathbf{0}$, we can discard the ﬁrst term on the right-hand side of the above expression, leading to the approximation
\begin{equation}
    \frac{\partial{\mathbf{r}'}}{\partial{\mathbf{\Deltaq}_t}}\approx \mathbf{\Phi}^T\mathbf{J}_t^T\mathbf{\Psi}_{\mathbf{J}}\mathbf{\Psi}_{\mathbf{J}}^T \mathbf{J}_t\mathbf{\Phi},
\end{equation}
which results in the following Newton-Raphson iterations: for $k=1,\hdots,K$, solve
\begin{equation}
\begin{aligned}
    \left[\mathbf{J}_{t}^{(k)}\tildeutkm \mathbf{\Phi}\right]^T\mathbf{\Psi}_{\mathbf{J}}\mathbf{\Psi}_{\mathbf{J}}^T\mathbf{J}_{t}^{(k)}\tildeutkm  \mathbf{\Phi}  \hatptk & = -\left[\mathbf{J}_{t}^{(k)}\tildeutkm  \mathbf{\Phi}\right]^T\mathbf{\Psi}_{\mathbf{J}}\mathbf{\Psi}_{\mathbf{J}}^T\mathbf{R}_t^{(k)}\tildeutkm \\
    \mathbf{\Deltaq}_t^{(k+1)} & =\mathbf{\Deltaq}_t^{(k)}+\alpha_t^{(k)}\hatptk\\
    \tildeu_t^{(k+1)}&=\tildeu_t^{(k)}+\mathbf{\Phi}\mathbf{\Deltaq}_t^{(k+1)}.
    \label{Newton-Raphson PG-Jacobian}
\end{aligned}
\end{equation}
It should be noted that solving the system of equations above amounts to minimizing the expression
\begin{equation}
    \min_{\hatpt\in\mathbb{R}^n} \norm\Big{\mathbf{\Psi}_{\mathbf{J}}^T\mathbf{J}_t\tildeutm \mathbf{\Phi} \hatpt-\mathbf{\Psi}_{\mathbf{J}}^T\mathbf{R}_t\tildeutm}^2
\end{equation}
at each iteration. This minimization problem can be addressed by using the QR factorization of 
\begin{equation}
    \mathbf{\Psi}_{\mathbf{J}}^T\mathbf{J}_t\mathbf{\Phi} \in \mathbb{R}^{m\times n}.
\end{equation}
Note that although this matrix is still rectangular, it has a much smaller dimension ($m<<N$ $\text{and}$ $n<<N$) compared to the full order model $\in \mathbb{R}^{N\times N}$.\\
\textbf{Proposition:}
\textit{If $\epsilon_{\mathbf{\Psi}_{\mathbf{J}}}=\mathbf{0}$ in the SVD of Eq.\ref{Jacobian SVD}, then Eq.\ref{Prime residual} is equivalent to Eq.\ref{Petrov Galerkin constrained Residual} for the training parameters}.\\ \\ 
\begin{proof}
\textit{Eq.\ref{Prime residual} can be written as}
\begin{equation}
    (\mathbf{\Psi}_{\mathbf{J}}\mathbf{\Psi}_{\mathbf{J}}^T\mathbf{J}_t\mathbf{\Phi})^T\mathbf{R}_t=\mathbf{0}.
\end{equation}
\textit{If $\epsilon_{\mathbf{\Psi}_{\mathbf{J}}}=\mathbf{0}$ in Eq.\ref{Jacobian SVD}, then $\mathbf{J}_t\mathbf{\Phi} \in col(\mathbf{\Psi}_{\mathbf{J}})$ for the training parameters. This means that, since $\mathbf{\Psi}_{\mathbf{J}}$ is orthogonal, $\mathbf{\Psi}_{\mathbf{J}}\mathbf{\Psi}_{\mathbf{J}}^T\mathbf{J}_t\mathbf{\Phi}=\mathbf{J}_t\mathbf{\Phi}$. Thus, Eq.\ref{Jacobian SVD} becomes}
\begin{equation}
    (\mathbf{J}_t\mathbf{\Phi})^T\mathbf{R}_t=\mathbf{0}.
\end{equation}
\end{proof}
\noindent As the number of modes increases, such as when solving highly nonlinear problems, the size of the snapshots matrix in Eq. \ref{Snapshots Matrix Jacobian} can potentially become quite large. This is because the size of the matrix increases linearly with the number of modes. This observation highlights the need to explore alternative techniques that can better handle large, evolving datasets. One such approach is based on residual evolution, which we will discuss in more detail below.
\subsubsection{Residual-based approach}
\label{Residual-based approach}
\begin{figure}[p]
\centering
\fbox{\includegraphics[width=1\linewidth]{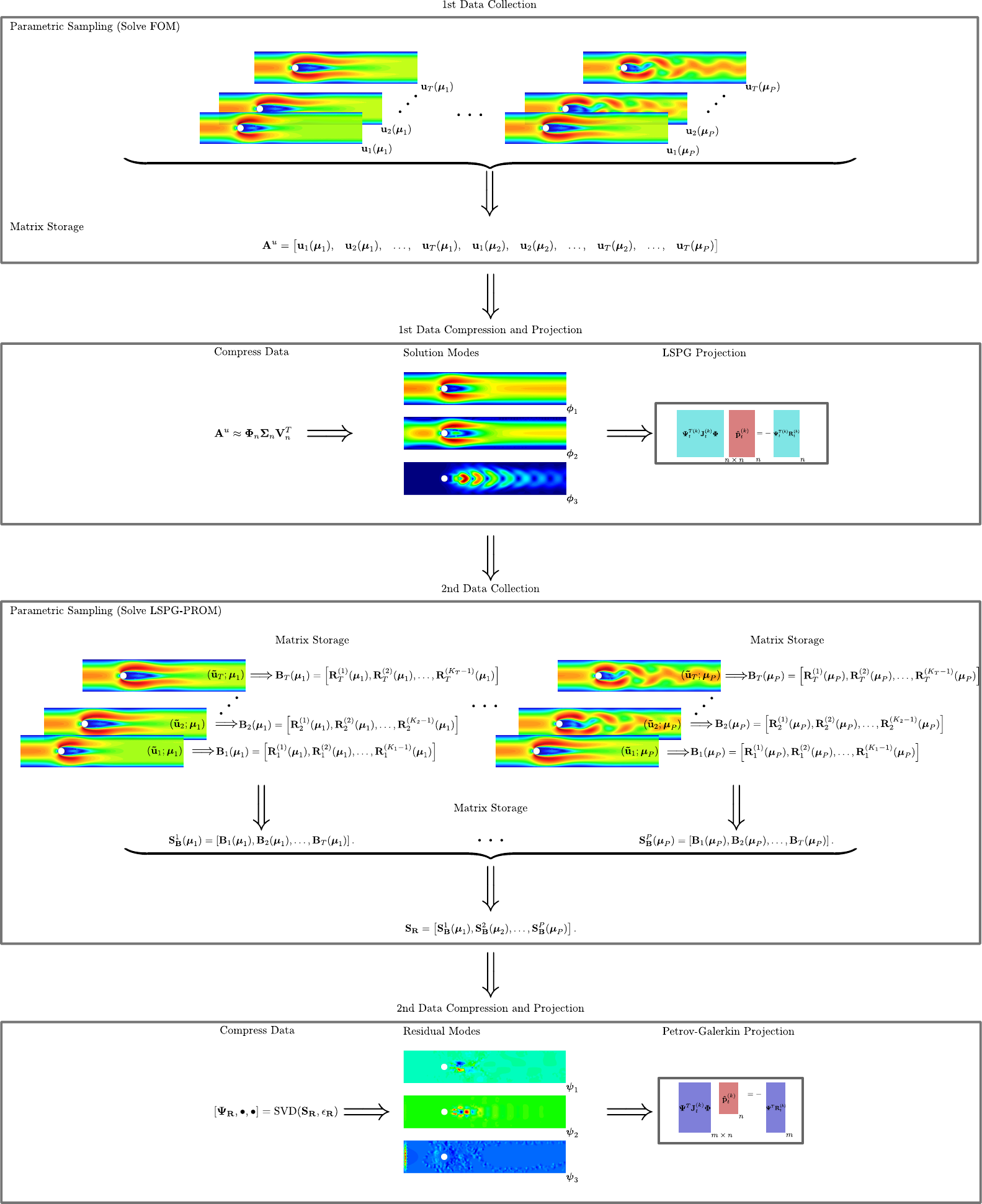}}
\caption{Schematic overview of Petrov-Galerkin PROM construction.}
\label{fig:Schematic overview of Petrov-Galerkin PROM construction.}
\end{figure}

The orthogonal matrix $\mathbf{\Psi}$ introduces an orthogonal decomposition of $\mathbb{R}^N$ as shown in Eq. \ref{Petrov Galerkin constrained Residual}:
\begin{equation}
    \mathbb{R}^N=null(\mathbf{\Psi}^T)\oplus col(\mathbf{\Psi}).
\end{equation}
It is important to note that a converged nodal residual vector will belong to $null(\mathbf{\Psi}^T)$, while non-converged nodal residual vectors, will have components in $col(\mathbf{\Psi})$. This observation suggests an alternative approach for determining the left ROB as a basis matrix for the column space of non-converged residuals. The detailed procedure for this approach is provided below.
\begin{framed}
\textit{Residuals basis procedure}
\begin{enumerate}
    \item For each training parameter $\boldsymbol{\mu}_j$ and at each nonlinear iteration $k$, define the non-converged residual at timestep $i$ as:
\begin{equation}
\mathbf{R}_i^{(k)}(\boldsymbol{\mu}_j)=\mathbf{R}_i(\mathbf{u}_{i}^{(k)};\boldsymbol{\mu}_j),
\end{equation}
Here, $k$ signifies the $k^{th}$ nonlinear iteration. For every timestep $i$ and each training parameter $\boldsymbol{\mu}_j$, gather a matrix of snapshots of the non-converged nodal residual vectors associated with the solutions $\tildeu_i^{(k)}$:

\begin{equation}
\mathbf{B}_{i}(\boldsymbol{\mu}_j) = \left[\mathbf{R}_i^{(1)}(\boldsymbol{\mu}_j), \mathbf{R}_i^{(2)}(\boldsymbol{\mu}_j), \hdots, \mathbf{R}_i^{(K_i-1)}(\boldsymbol{\mu}_j)\right],
\end{equation}
where $K_i$ denotes the total amount of non-linear iterations for the $i^{th}$ time step for the corresponding parameter $\boldsymbol{\mu}_j$.

\item Assemble all the snapshot matrices $\mathbf{B}_i(\boldsymbol{\mu}_j)$ into a singular snapshot matrix $\mathbf{S}^j_{\mathbf{B}}(\boldsymbol{\mu}_j)$ for each parameter $\boldsymbol{\mu}_j$. Specifically, stack the snapshot matrices $\mathbf{B}_i(\boldsymbol{\mu}_j)$ for all timesteps and the $j^{th}$ parameter $\boldsymbol{\mu}_j$ into a singular matrix representing all non-converged residuals:

\begin{equation}
\mathbf{S}^j_{\mathbf{B}}(\boldsymbol{\mu}_j) = \left[\mathbf{B}_1(\boldsymbol{\mu}_j),\mathbf{B}_2(\boldsymbol{\mu}_j),\hdots, \mathbf{B}_T(\boldsymbol{\mu}_j)\right].
\end{equation}

\item Compile all the snapshot matrices $\mathbf{S}^j_{\mathbf{B}}(\boldsymbol{\mu}_j)$ into a final non-converged residuals matrix for all training parameters:

\begin{equation}
\mathbf{S}_{\mathbf{R}} = \left[\mathbf{S}^1_{\mathbf{B}}(\boldsymbol{\mu}_1),\mathbf{S}^2_{\mathbf{B}}(\boldsymbol{\mu}_2),\hdots, \mathbf{S}^P_{\mathbf{B}}(\boldsymbol{\mu}_P)\right].
\end{equation}
\item Obtain the orthogonal basis $\mathbf{\Psi}_{\mathbf{R}}$:\\ \\
Perform a truncated SVD to obtain the orthogonal basis for their column spaces:
\begin{equation}
    \left[\mathbf{\Psi}_{\mathbf{R}},\bullet,\bullet\right] = \text{SVD}(\mathbf{S}_\mathbf{R},\epsilon_{\mathbf{R}}).
\end{equation}
It is clear that
\begin{equation}
    col(\mathbf{\Psi}_{\mathbf{R}})\subseteq col(\mathbf{\Psi}_{\mathbf{J}})
\end{equation}
(see Fig.\ref{fig:Schematic overview of Petrov-Galerkin PROM construction.} for schematic overview)
 
\end{enumerate}
\end{framed}
It is worth mentioning that, for highly nonlinear dynamical systems, the number of modes in the right ROB is expected to be higher than the number of non-linear iterations. This implies that the size of the $\mathbf{S}_{\mathbf{R}}$ matrix, which is constructed by gathering non-converged residuals for each parameter, will only increase in proportion to the number of non-linear iterations and is expected to be smaller than the size of the $\mathbf{S}_{\mathbf{J}}$ matrix, which is constructed using the Jacobian matrix projected onto the left ROB.

\subsection{Comparing Matrix Sizes and Computational Costs in Iterative Procedures: The Case of Rectangular Petrov-Galerkin Approach}

Reduced-order modeling involves iterative procedures, for example, in the Newton-Raphson method, the matrix size plays a crucial role in determining both the computational cost and the solution approach.

\begin{figure}[p]
\centering
\fbox{\includegraphics[width=1\linewidth]{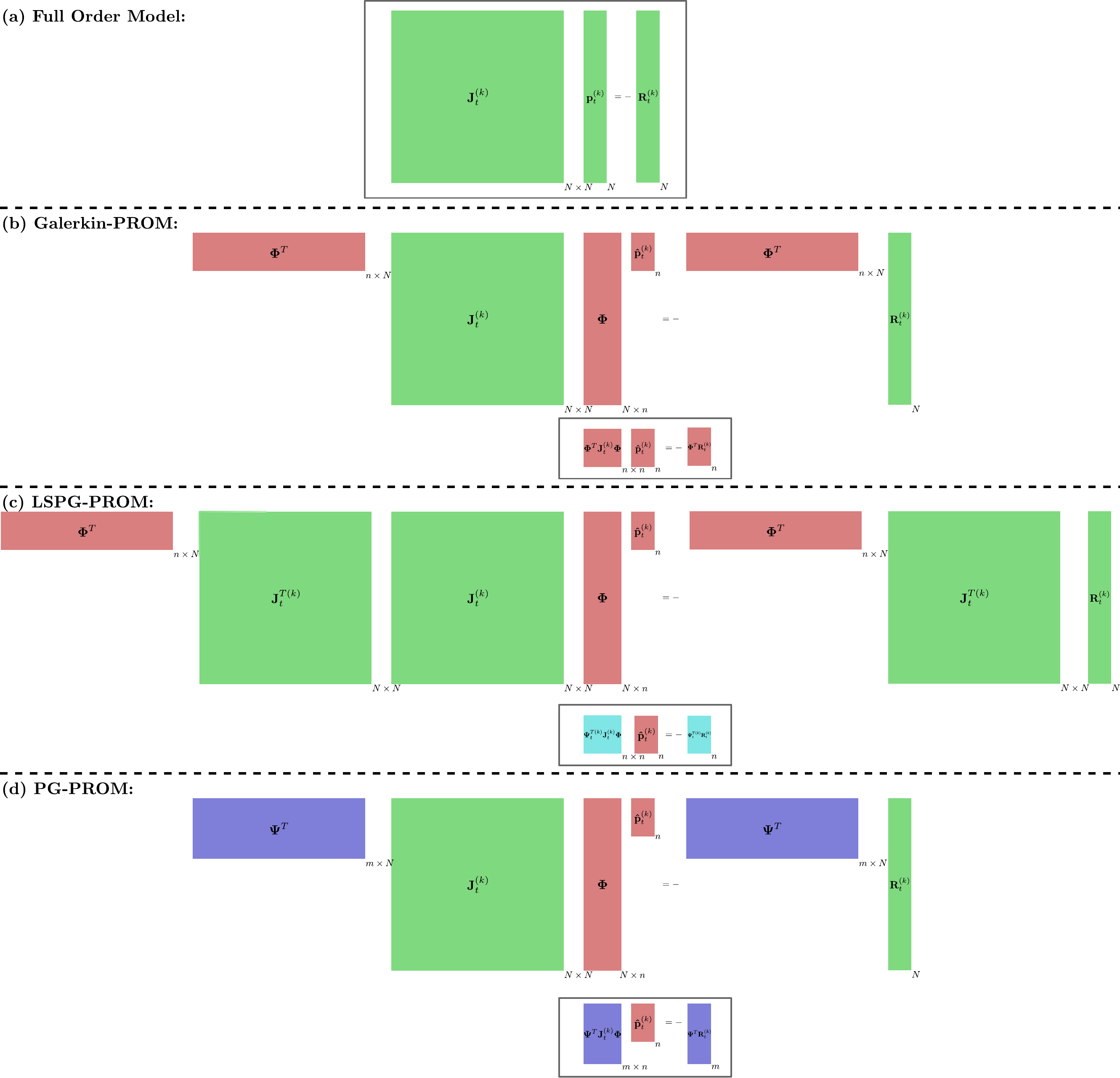}}
\caption{This diagram illustrates the various dimensions of the systems that need to be addressed in different scenarios, as well as their corresponding projections. The four cases depicted are: (a) a full-order model, (b) a Galerkin-PROM, (c) a LPSG-PROM, and (d) a PG-PROM.}
\label{fig:Systems' size.}
\end{figure}

Let us consider the matrix sizes involved in three different iterative schemes: the FOM, Galerkin ROM, and Petrov-Galerkin ROM. The FOM system involves a matrix 
\begin{equation*}
    \mathbf{J}_t\in \mathbb{R}^{N\times N},
\end{equation*}
where $N$ is the original dimensionality of the system. The Galerkin projection, on the other hand, employs the matrix 
\begin{equation*}
    \mathbf{\Phi}^T\mathbf{J}_t\mathbf{\Phi}\in \mathbb{R}^{n\times n},
\end{equation*} 
which results in a smaller system ($n<<N$) that needs to be solved, leading to significant computational savings during the iterative procedure. It is worth noting that the full order model, Galerkin ROM, and LSPG ROM systems considered thus far have all been square systems. However, the Petrov-Galerkin projection requires solving a rectangular system of size 
\begin{equation*}
    \mathbf{\Psi}^T\mathbf{J}_t\mathbf{\Phi}\in \mathbb{R}^{m\times n},
\end{equation*} 
where $m\geq n$ (with $\mathcal{O}(m)=\mathcal{O}(n)$), leading to a minimization problem that needs to be addressed. One possible solution to this problem is QR factorization. Figure \ref{fig:Systems' size.} illustrates a schematic representation of the various systems and their resulting sizes. Additionally, Algorithm \ref{alg:Newton PG-invariant left ROB} presents the invariant left ROB PG-PROM algorithm, which differs from the general PG-PROM algorithm in Algorithm \ref{alg:Newton PG} as the left ROB $\mathbf{\Psi}$ is an input rather than an iteration-dependent variable. This feature brings significant computational savings in the subsequent hyper-reduction stage, as we will see later on. Despite the increased system size and associated computational cost, the Petrov-Galerkin projection can provide better accuracy than the Galerkin projection in certain situations.

\noindent 
Let us examine the computational costs tied to the training and online phases of the Galerkin, LSPG, and Petrov-Galerkin strategies more closely. Both the Galerkin and LSPG strategies involve a FOM assembly and solution for parameters $P$ and their respective times $T$ during the training phase. The complexity of the FOM assembly is $\mathcal{O}(L)$, and for the FOM solution, it is $\mathcal{O}(N^k)$, with $k\geq 1$ depending on the specific iterative solver employed for the sparse system. Then, a Singular Value Decomposition (SVD) compression is performed with a complexity of $\mathcal{O}(N(TP)n)$ using a randomized SVD. In the subsequent online phase, the FOM assembly is repeated, followed by a system projection operation, with complexities of $\mathcal{O}(N_0 \times n)$ and $\mathcal{O}(n^2 \times N)$ for sparse-dense and dense-dense matrix multiplication respectively, where $N_0$ represents the non-zero entries. The online phase concludes with a direct solution of a dense ROM, which is associated with a complexity of $\mathcal{O}(n^3)$. Conversely, the Petrov-Galerkin strategy introduces an additional training phase, which involves a repeated FOM assembly, system projection, and solution of the LSPG ROM for the same parameters $P$ and their corresponding times $T$. This phase also requires an additional SVD compression operation, with a complexity of $\mathcal{O}(N(TPK)n)$, reflecting the sum of nonlinear iterations ($K$) in a residual-based approach, and once again employing a randomized SVD. Despite these heightened computational costs, this phase bolsters the HROM scheme by negating the need for a complementary mesh during the creation of the HROM model. The online phase mirrors that of the Galerkin and LSPG strategies but adjusts the ROM solution complexity to $\mathcal{O}(mn^2)$ (with $\mathcal{O}(m)=\mathcal{O}(n)$).

\begin{algorithm}
\caption{Newton-Raphson strategy for invariant left ROB PG-PROM.}
\label{alg:Newton PG-invariant left ROB}
\hspace*{\algorithmicindent} \textbf{Input:} State variable $\uref$, right  ROB $\mathbf{\Phi}$, and left ROB $\mathbf{\Psi}$\\
\hspace*{\algorithmicindent} \textbf{Output:} State variable $\tildeu_t$ and increment update $\deltauhat_t$
\begin{algorithmic}[1]
    \STATE Initialize $\mathbf{\Deltaq}_t^{(0)}=\mathbf{0}$, i.e., $\tildeu_t^{(0)}=\uref$
    \FOR{$k=0,\hdots,K $ (convergence)}
        \STATE Evaluate residual $\mathbf{R}_t^{(k)}=\left.\mathbf{R}_t\right|_{\tildeu_t^{(k)}}$
        \STATE Evaluate Jacobian $\mathbf{J}_t^{(k)}=\left.\frac{\partial \mathbf{R}_t}{\partial \mathbf{\tildeu}}\right|_{\tildeu_t^{(k)}}$
        \STATE Compute $\mathbf{W}^{(k)} = \mathbf{\Psi}^T\mathbf{J}_{t}^{(k)}\mathbf{\Phi}$
        \IF{$m\neq n$}
            \STATE Compute QR decomposition $\mathbf{W}^{(k)} = \mathbf{Q}^{(k)} \mathbf{D}^{(k)}$
            \STATE Solve $\mathbf{D}^{(k)}  \hatptk = -\mathbf{Q}^{T(k)}\mathbf{\Psi}^T\mathbf{R}_t^{(k)}$
        \ELSE
            \STATE Solve $\mathbf{W}^{(k)}  \hatptk = -\mathbf{\Psi}^T\mathbf{R}_t^{(k)}$
        \ENDIF
        % \IF{line-search}
        %     \STATE Compute $\alpha_t^{(k)}$
        % \ELSE
        %     \STATE $\alpha_t^{(k)} = 1$
        % \ENDIF
        \STATE Update $\mathbf{\Deltaq}_t^{(k+1)} =\mathbf{\Deltaq}_t^{(k)}+\alpha_t^{(k)}\hatptk$
        \STATE Update $\tildeu_t^{(k+1)}=\tildeu_t^{(k)}+\mathbf{\Phi}\mathbf{\Deltaq}_t^{(k+1)}$
    \ENDFOR
    \STATE $\deltauhat_t=\mathbf{\Phi}\sum_{k=1}^K\mathbf{\Deltaq}_t^{k+1}$
    \STATE $\tildeu_t=\tildeu_t^{(K+1)}\equiv \uref+\deltauhat_t$
\end{algorithmic}
\end{algorithm}
\subsection{Hyper-reduction}
The Newton-Raphson iterations for the invariant left ROB Petrov-Galerkin Projection-based Reduced Order Model with residual-based formulation, aforementioned: for $k=1,\hdots,K$, we solve the following equations:\\
\begin{equation}
\boxed{\begin{aligned} \mathbf{\Psi}_{\mathbf{R}}^T\mathbf{J}_{t}^{(k)}\tildeutkm \mathbf{\Phi}  \hatptk & = -\mathbf{\Psi}_{\mathbf{R}}^T\mathbf{R}_t^{(k)}\tildeutkm\\
\mathbf{\Deltaq}_t^{(k+1)} & =\mathbf{\Deltaq}_t^{(k)}+\alpha_t^{(k)}\hatptk\\
\tildeu_t^{(k+1)}&=\tildeu_t^{(k)}+\mathbf{\Phi}\mathbf{\Deltaq}_t^{(k+1)}.
\label{Newton-Raphson PG-invariant left ROB}
\end{aligned}}
\end{equation}
We can highlight the difference between the variant and invariant left ROB approaches by revisiting Eq. \ref{Galerkin assembly of elemental contributions}, which, when substituted with the residual-based left ROB, gives:

\begin{equation}
\begin{aligned}
\mathbf{\Psi}_{\mathbf{R}}^T\mathbf{R}_t&=\mathbf{\Psi}_{\mathbf{R}}^T\sum_{e=1}^{L}\mathbf{L}^{e^T}\mathbf{R}_t^{e}\\
&=\sum_{e=1}^{L}(\mathbf{L}^{e}\mathbf{\Psi}_{\mathbf{R}})^T\mathbf{R}_t^{e}\\
&=\boxed{\sum_{e=1}^{L}\mathbf{\Psi}_{\mathbf{R}}^{e^T}\mathbf{R}_t^{e}}.\\
\end{aligned}
\end{equation} 

Comparing this to Eq. \ref{Leasts Squares Projected Residuals}, the main difference in the LSPG assembling in the variant approach is the term $\barrte$ (Eq. \ref{Elemental Residual multiplied by the elemental jacobian}). This term requires determining residual vectors for all elements that share nodes with element $e$ (i.e., a patch of elements), which leads to the need to store a complementary mesh for the online evaluation of the LSPG-HPROM. In contrast, the left ROB in the PG-HPROM method is invariant and does not depend on surrounding elements. Therefore, the hyper-reduction scheme can be efficiently coupled with the ECM algorithm using an element-by-element construction and optimal mesh sampling.

\section{Results}
\begin{itemize}
\item Analysis of the nonlinear behavior of a cantilever beam subjected to prescribed forces in structural mechanics, with hyperelastic material properties described by the Kirchhoff Saint-Venant model.
\item A transient rotating pulse convection-diffusion problem with convection dominance.
\end{itemize}
To evaluate the accuracy of the constructed PROMs and HPROMs, we measure the relative error between the solution state variables and their corresponding values obtained from the FOM and PROM-based models. At each $i^{th}$ time-step solution snapshot of the $j^{th}$ parameter configuration, the relative error is calculated as:
\begin{equation}
E_{i}(\mu_j) = \frac{\norm\Big{\tilde{\mathcal{\mathbf{u}}}_i(\mu_j) -\mathcal{\mathbf{u}}_i(\mu_j) }^2}{\norm\Big{\mathcal{\mathbf{u}}_i(\mu_j) }^2},
\end{equation}
\noindent where $\tilde{\mathcal{\mathbf{u}}}_i(\mu_j)\in\mathbb{R}^N$ and $\mathcal{\mathbf{u}}_i(\mu_j)\in\mathbb{R}^N$ are the approximated and reference solutions of the state variable, respectively, at the $i^{th}$ time step for the $j^{th}$ parameter configuration. We also calculate the relative error for the entire set of solution snapshots, which is defined as:
\begin{equation}
E = \frac{\norm\Big{\tilde{\mathcal{\boldsymbol{S}}}_{\mathbf{u}}-\mathcal{\boldsymbol{S}}_{\mathbf{u}}}_F}{\norm\Big{\mathcal{\boldsymbol{S}}_{\mathbf{u}}}_F},
\end{equation}
where
\begin{equation}
\tilde{\mathcal{\boldsymbol{S}}}_{\mathbf{u}}=
\begin{bmatrix}
\tilde{\mathbf{u}}_1(\boldsymbol{\mu}_1),&\tilde{\mathbf{u}}_2(\boldsymbol{\mu}_1), &\hdots, &\tilde{\mathbf{u}}_T(\boldsymbol{\mu}_1),&\tilde{\mathbf{u}}_1(\boldsymbol{\mu}_2), &\hdots, &\tilde{\mathbf{u}}_T(\boldsymbol{\mu}_2), &\hdots, &\tilde{\mathbf{u}}_T(\boldsymbol{\mu}_P)]
\end{bmatrix}
\end{equation}
and
\begin{equation}
\mathcal{\boldsymbol{S}}_{\mathbf{u}}=
\begin{bmatrix}
\mathbf{u}_1(\boldsymbol{\mu}_1), &\mathbf{u}_2(\boldsymbol{\mu}_1),&\hdots,&\mathbf{u}_T(\boldsymbol{\mu}_1),&\mathbf{u}_1(\boldsymbol{\mu}_2), &\mathbf{u}_2(\boldsymbol{\mu}_2),&\hdots,& \mathbf{u}_T(\boldsymbol{\mu}_2),& \hdots,&\mathbf{u}_T(\boldsymbol{\mu}_P)
\end{bmatrix}
\end{equation}
are the snapshot matrices capturing the solution and approximation state variables across all time steps and parameter configurations.

\subsection{Structural Mechanics Case}
\begin{figure}[h!]
\centering
\fbox{\includegraphics[width=0.5\linewidth]{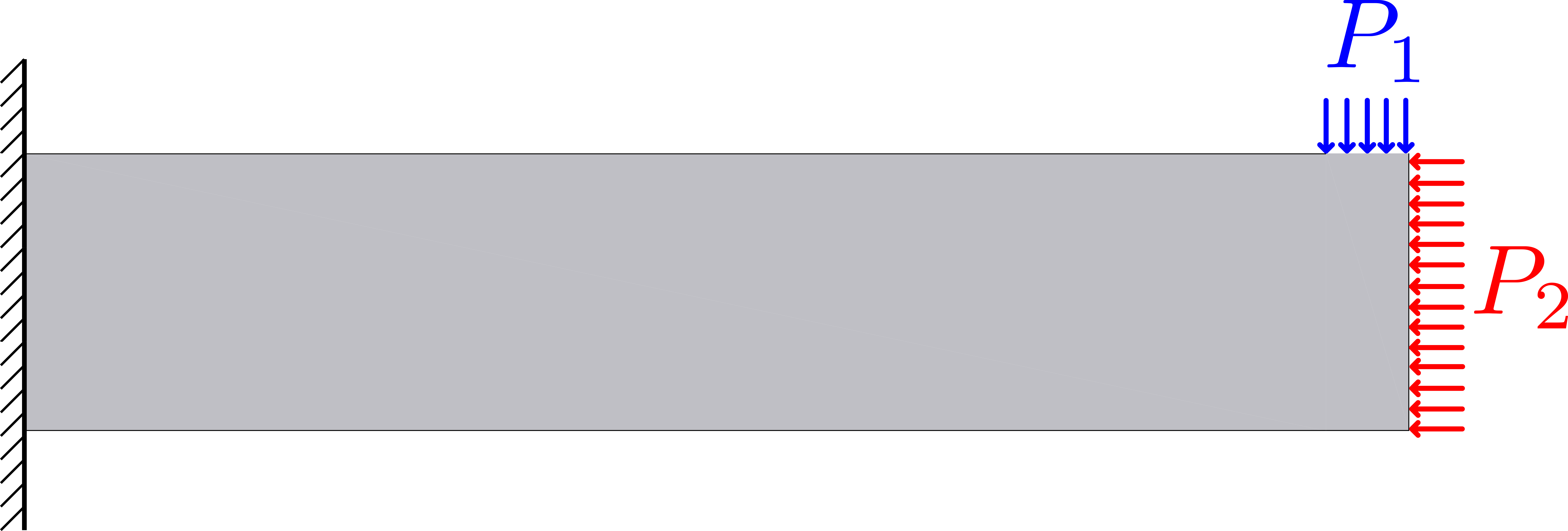}}
\caption{Cantilever beam with two pressure loads.}
\label{fig:Nonlinear cantilever beam diagram}
\end{figure}

In this research, we analyze the nonlinear behavior of a cantilever beam in the field of structural mechanics, with the primary objective to compare the performance of Galerkin-PROMs, LSPG-PROMs, and PG-PROMs for systems governed by SPD operators. Additionally, the Jacobian-based and Residual-based approaches for selecting the left reduced order basis (ROB) $\boldsymbol{\Psi}$ are compared. The cantilever beam, subjected to prescribed forces, is assumed to possess hyperelastic material properties, as described by the Kirchhoff Saint-Venant model. The material properties include a Young's modulus of 206.9 GPa and a Poisson's ratio of 0.29. This beam is designed to undergo large displacements, with the problem being formulated using a Total Lagrangian formulation under a plane stress analysis setup. Two pressure loads, $P_1$ and $P_2$, are applied to the system. They are defined as $P_1=c_1*\sqrt{\alpha}$ and $P_2=c_2*\sqrt{\alpha}$ respectively, with $c_1=10^8$ Newtons and $c_2=10^7$ Newtons. These loads, directed in negative vertical and horizontal directions, are presented in Fig.\ref{fig:Nonlinear cantilever beam diagram}. The loads are applied simultaneously with varying intensities over a series of steps to simulate the nonlinear behavior of the cantilever beam. The model will be trained for $\alpha$ values in the range of $[0.1, 2.0]$ and tested for extrapolated values of $\alpha$ in the range $(2.0, 4.0]$. As also depicted in Fig.\ref{fig:Nonlinear cantilever beam diagram}, there is a fixed displacement imposed on the left boundary. The model for this analysis comprises 6954 finite elements with 3626 nodes, leading to 7252 degrees of freedom. This comprehensive model complexity aims to provide a precise simulation of the nonlinear behavior of the system under the conditions specified.
\paragraph{Reduction}
For this analysis, specific tolerances were set to guide the reduction process. The tolerance used to determine the dimensionality of the latent variables $\mathbf{\hat{u}}$, which subsequently defines the number of columns in the right reduced order basis (ROB) $\mathbf{\Phi}$, was set at $\epsilon_u= 10^{-6}$. For the Petrov-Galerkin approaches, both Residual-based and Jacobian-based, the singular value decomposition (SVD) tolerance was established as $\epsilon_{\mathbf{R}}=\epsilon_{\mathbf{J}}= 10^{-6}$. This tolerance value serves to determine the size of the left ROB $\mathbf{\Psi}$. Lastly, the tolerance for the empirical cubature method (ECM), used across all projection strategies, was set to machine precision.
\paragraph{Discussion}
The numerical investigation led to a latent approximation space $\boldsymbol{\Phi}$ of dimension 3 (number of modes) and a constraint space $\boldsymbol{\Psi}$ of dimension 7 (number of modes). The latter size was consistent in both Petrov-Galerkin methods, Jacobian-based and Residual-based, highlighting the uniformity in the constraint space selection.
\begin{table}[h!]
\centering
\caption{Overall L2 Norms for Training Phase}
\begin{tabular}{|c|c|c|c|c|c|}
\hline
 & \multicolumn{3}{c|}{Overall L2 Norm (Training)} & \\
\cline{2-4}
\multirow{1}{*}{\centering Strategy}& \multicolumn{1}{c|}{FOM vs ROM} & \multicolumn{1}{c|}{ROM vs HROM} & \multicolumn{1}{c|}{FOM vs HROM} & \multicolumn{1}{c|}{Variable}\\
\cline{1-5}
& $2.125 \times 10^{-8}$ & $9.630 \times 10^{-13}$ & $2.125 \times 10^{-8}$ & Displacement $x$ \\
\cline{2-5}
Galerkin & $4.089 \times 10^{-9}$ & $1.060 \times 10^{-12}$ & $4.089 \times 10^{-9}$ & Displacement $y$ \\
\hline
& $1.257 \times 10^{-6}$ & $1.251 \times 10^{-9}$ & $1.257 \times 10^{-6}$ & Displacement $x$ \\
\cline{2-5}
LSPG & $1.136 \times 10^{-6}$ & $1.173 \times 10^{-9}$ & $1.136 \times 10^{-6}$ & Displacement $y$ \\
\hline
& $9.731 \times 10^{-7}$ & $1.434 \times 10^{-8}$ & $9.587 \times 10^{-7}$ & Displacement $x$ \\
\cline{2-5}
Petrov-Galerkin Res. & $8.826 \times 10^{-7}$ & $1.325 \times 10^{-8}$ & $8.693 \times 10^{-7}$ & Displacement $y$ \\
\hline
& $9.060 \times 10^{-7}$ & $1.159 \times 10^{-8}$ & $8.946 \times 10^{-7}$ & Displacement $x$ \\
\cline{2-5}
Petrov-Galerkin Jac. & $7.814 \times 10^{-7}$ & $1.021 \times 10^{-8}$ & $7.712 \times 10^{-7}$ & Displacement $y$ \\
\hline
\end{tabular}
\label{StructuralTrainL2}
\end{table}

\begin{figure}[h]
    \centering
    \fbox{\includegraphics[width=\textwidth]{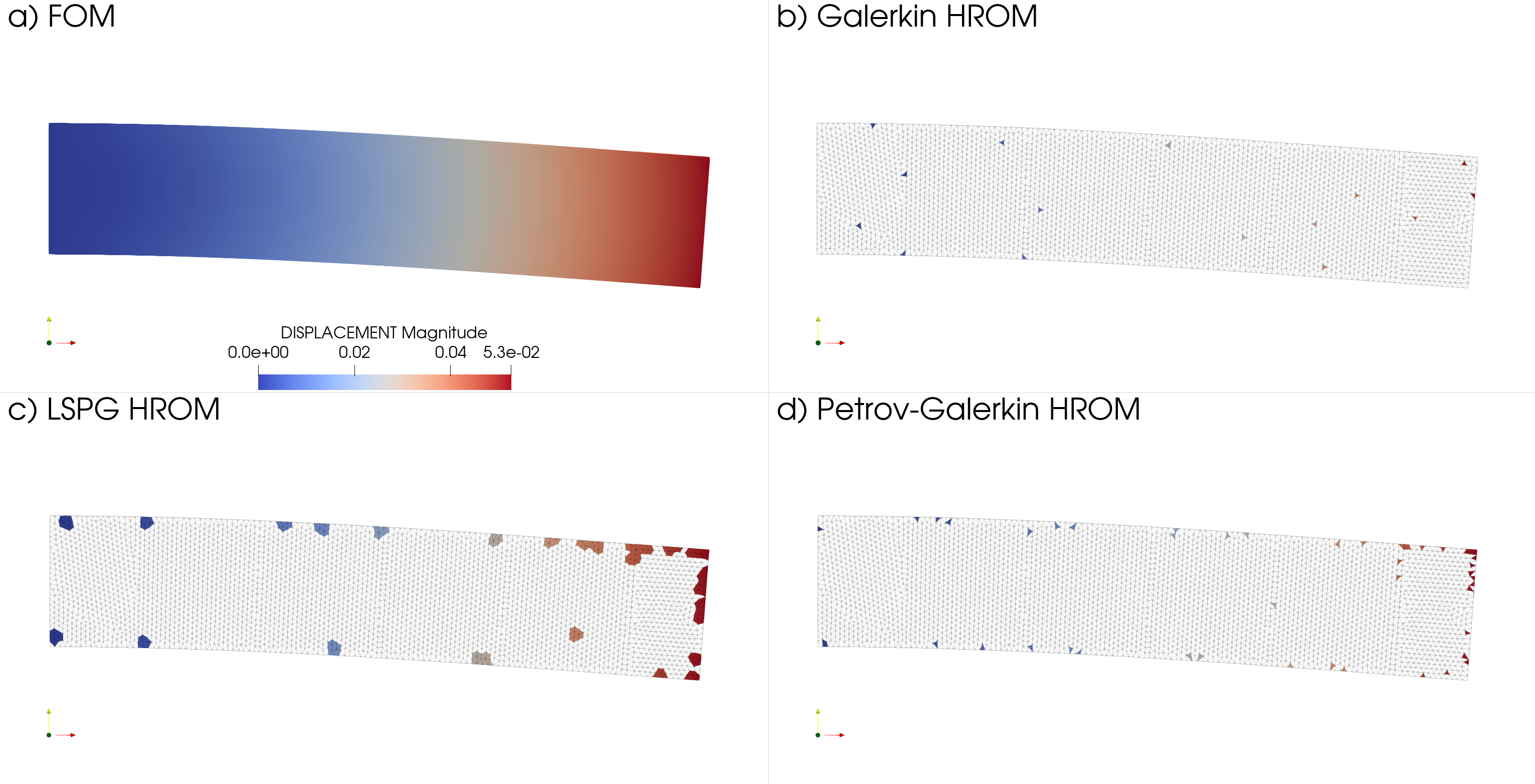}}
    \caption{Comparison of HROM solutions for different strategies in the context of the cantilever beam: (a) FOM, (b) Galerkin HROM, (c) LSPG HROM, (d) Petrov-Galerkin HROM. The figure also exhibits the different HROM meshes used for each strategy.}
    \label{fig:structural_hrom_strategies}
\end{figure}

\begin{table}[h!]
\centering
\caption{Overall L2 Norms for Testing Phase}
\begin{tabular}{|c|c|c|c|c|c|}
\hline
 & \multicolumn{3}{c|}{Overall L2 Norm (Testing)} & \\
\cline{2-4}
\multirow{1}{*}{Strategy} & \multicolumn{1}{c|}{FOM vs ROM} & \multicolumn{1}{c|}{ROM vs HROM} & \multicolumn{1}{c|}{FOM vs HROM} & \multicolumn{1}{c|}{Variable}\\
\cline{1-5}
& $3.331 \times 10^{-7}$ & $2.748 \times 10^{-11}$ & $3.331 \times 10^{-7}$ & Displacement $x$ \\
\cline{2-5}
Galerkin & $6.843 \times 10^{-8}$ & $3.326 \times 10^{-11}$ & $6.842 \times 10^{-8}$ & Displacement $y$ \\
\hline
& $2.095 \times 10^{-5}$ & $1.230 \times 10^{-8}$ & $2.094 \times 10^{-5}$ & Displacement $x$ \\
\cline{2-5}
LSPG & $1.968 \times 10^{-5}$ & $1.178 \times 10^{-8}$ & $1.967 \times 10^{-5}$ & Displacement $y$ \\
\hline
& $1.618 \times 10^{-5}$ & $2.339 \times 10^{-7}$ & $1.595 \times 10^{-5}$ & Displacement $x$ \\
\cline{2-5}
Petrov-Galerkin Res. & $1.528 \times 10^{-5}$ & $2.243 \times 10^{-7}$ & $1.506 \times 10^{-5}$ & Displacement $y$ \\
\hline
& $1.508 \times 10^{-5}$ & $2.279 \times 10^{-7}$ & $1.486 \times 10^{-5}$ & Displacement $x$ \\
\cline{2-5}
Petrov-Galerkin Jac. & $1.354 \times 10^{-5}$ & $2.020 \times 10^{-7}$ & $1.334 \times 10^{-5}$ & Displacement $y$ \\
\hline
\end{tabular}
\label{StructuralTestL2}
\end{table}
\noindent Analyzing the overall L2 norms in both the training and testing phases, our models have delivered highly promising results. In the training phase (Table \ref{StructuralTrainL2}), the L2 norms between FOM, ROM, and HROM models were found to be on the order of $10^{-6}$ to $10^{-8}$ for both x and y displacements. This corresponds closely to our original error truncation tolerance set at $10^{-6}$, thereby demonstrating excellent model fidelity in reproducing the training scenarios. Moreover, both the LSPG and Petrov-Galerkin strategies yielded similarly low L2 norms, reinforcing the residual-minimum property these methods aim to achieve. This result not only validates our proposed methodologies, but also highlights the underlying relationship between these projection strategies.  In the testing phase (Table \ref{StructuralTestL2}), where the models are applied to an extrapolated parametric space, the L2 norms remained low and within the same order of magnitude as the original error truncation tolerance. This result validates the robustness of our models, as they maintain a high degree of accuracy even when extrapolated beyond their training data.\\ \\
The HROMs selection process resulted in the following distribution of elements: Galerkin strategy yielded 20 elements (0.29\% of the total 6954 elements), LSPG strategy yielded 35 elements (0.50\% of the total), Petrov-Galerkin Residual-based strategy selected 68 elements (0.98\% of the total), and Petrov-Galerkin Jacobian-based strategy selected 54 elements (0.78\% of the total). It is worth noting that although the LSPG strategy initially appears to require fewer selected elements than the Petrov-Galerkin strategies, it demands information from the surrounding elements and thus necessitates a complementary mesh. This increases the effective number of elements from 35 to 293, as shown in Figure \ref{fig:structural_hrom_strategies}. Consequently, despite the initially smaller selection, LSPG yields a considerably lower speedup when compared to the other strategies.

\begin{table}[h!]
\begin{center}
\begin{tabular}{|c|c|}
\hline
Strategy & Total Speedup\\
\hline
ROM Galerkin & 9.92\\
HROM Galerkin & 612.13\\
ROM LSPG & 13.32\\
\cellcolor{gray!25}HROM LSPG & \cellcolor{gray!25}119.85\\
ROM Petrov-Galerkin Res. & 7.70\\
HROM Petrov-Galerkin Res. & 164.70\\
ROM Petrov-Galerkin Jac. & 8.56\\
\cellcolor{gray!25}HROM Petrov-Galerkin Jac. & \cellcolor{gray!25}228.86\\
\hline
\end{tabular}
\caption{Speedups for ROM and HROM using different strategies.}
\label{tab:speedup}
\end{center}
\end{table}

\noindent These results elucidate the significant speedups achieved when deploying HROM strategies compared to their traditional ROM counterparts.\\ \\
Our numerical results highlight that, particularly for problems exhibiting symmetric positive-definite (SPD) operators like the one studied in this case, the Galerkin approach consistently provides superior results and speed-ups. However, the reader must not overlook the primary goal of this work, which is to propose an equivalent projection strategy to LSPG, one that utilizes a fixed left reduced order basis, $\boldsymbol{\Psi}$, and obviates the necessity of the complementary mesh.\\
Noteworthy is the contrast in the speed-ups achieved: we observe a transition from an LSPG HROM speed-up of 119.85 to a Petrov-Galerkin speed-up of 164.70 for the Residuals-based approach and 228.86 for the Jacobian-based method. This significant impact in terms of computational efficiency is pronounced even for this relatively simple and coarse problem. Another important aspect to discuss is the relative effort required to train the HROM models from the FOM. In this specific case, if we consider the effort for training the Galerkin strategy as a reference, we observe that the LSPG strategy has a slightly higher ratio of 1.01 to the Galerkin, showing a similar efficiency in training the model. On the other hand, the Petrov-Galerkin strategies present a higher ratio of 1.38. This increase in training effort for the Petrov-Galerkin strategies originates from the second training phase, which is required to determine the optimal fixed left ROB $\boldsymbol{\Psi}$. This underscores the trade-off in computational efficiency when selecting a modeling strategy. In this case, the reduction in elements needed for the Petrov-Galerkin methods (which eliminates the need for a complementary mesh, unlike LSPG) comes at the cost of increased computational effort during training.\\ \\
Moving forward, the aim is to investigate how these findings extrapolate to problems with finer meshes and different physics. We believe the strategies presented in this work will be invaluable tools in addressing more complex, real-world problems in structural mechanics and beyond.

\begin{figure}[h!]
    \centering
    \begin{subfigure}{.5\textwidth}
        \centering
        \fbox{\includegraphics[width=0.9\linewidth]{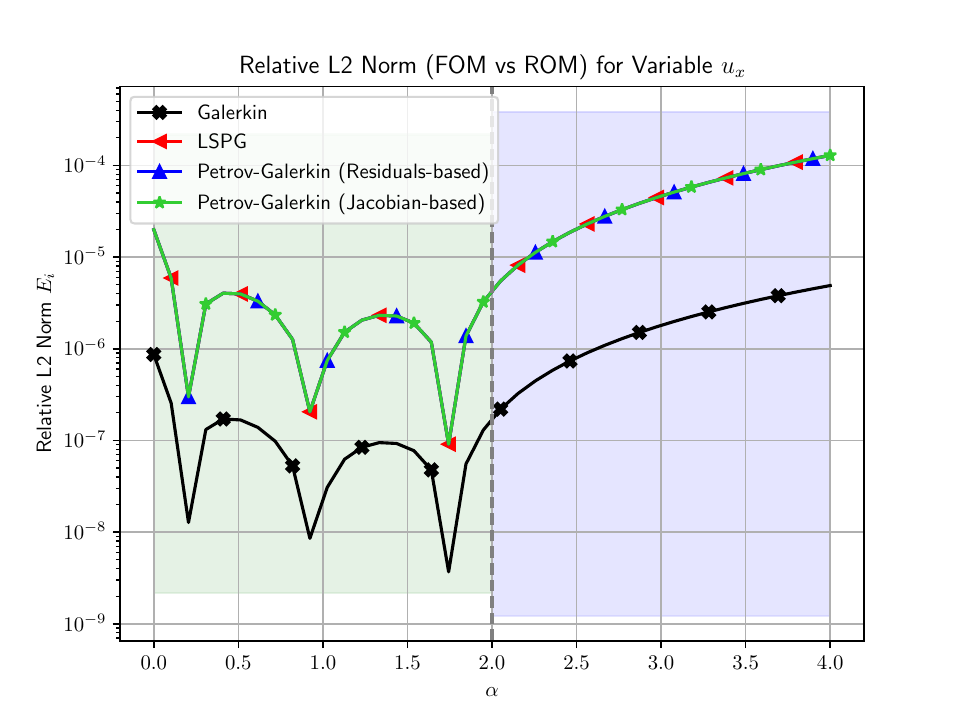}}
        \caption{Displacement in x}
        \label{fig:RelativeL2Norm_FOMvsROM_DISPLACEMENT_X}
    \end{subfigure}%
    \begin{subfigure}{.5\textwidth}
        \centering
        \fbox{\includegraphics[width=0.9\linewidth]{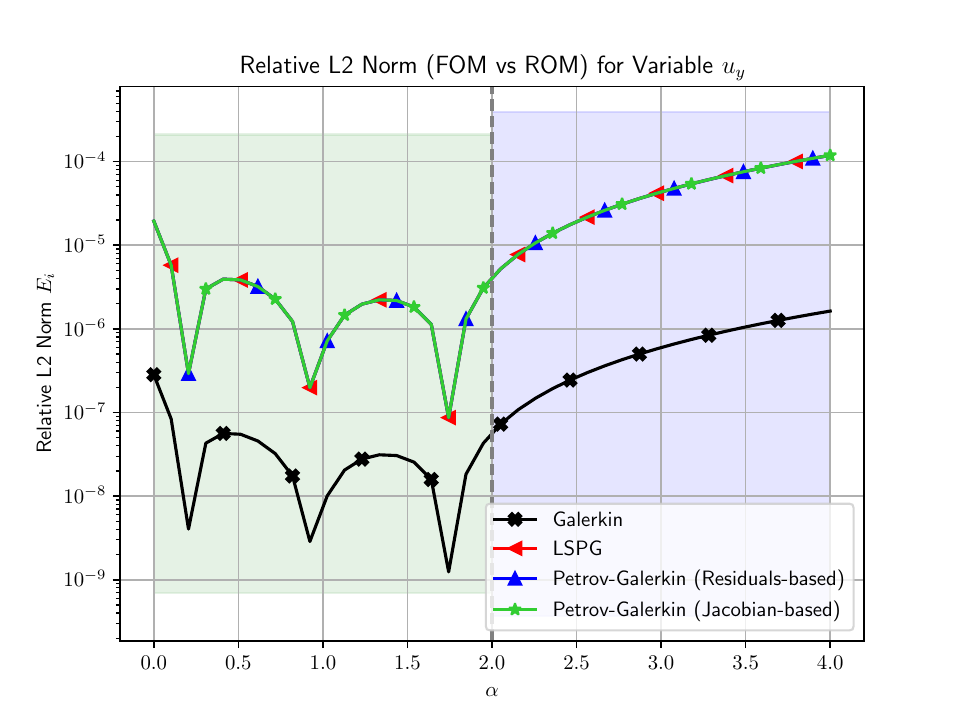}}
        \caption{Displacement in y}
        \label{fig:RelativeL2Norm_FOMvsROM_DISPLACEMENT_Y}
    \end{subfigure}
    \caption{Relative L2 Norm between the FOM and ROM against the scaling parameter $\alpha$. The vertical line at $\alpha$ = 2.0 denotes the boundary between the training ($\alpha$ $\in$ [0.0, 2.0], shaded green) and testing phases ($\alpha$ $\in$ (2.0, 4.0], shaded blue). Each line represents a different strategy used in the model reduction process. The subfigures represent two distinct variables of interest: (a) Displacement in x, and (b) Displacement in y.}
    \label{fig:RelativeL2Norm_FOMvsROM_DISPLACEMENT}
\end{figure}
\begin{figure}[h!]
    \centering
    \begin{subfigure}{.5\textwidth}
        \centering
        \fbox{\includegraphics[width=0.9\linewidth]{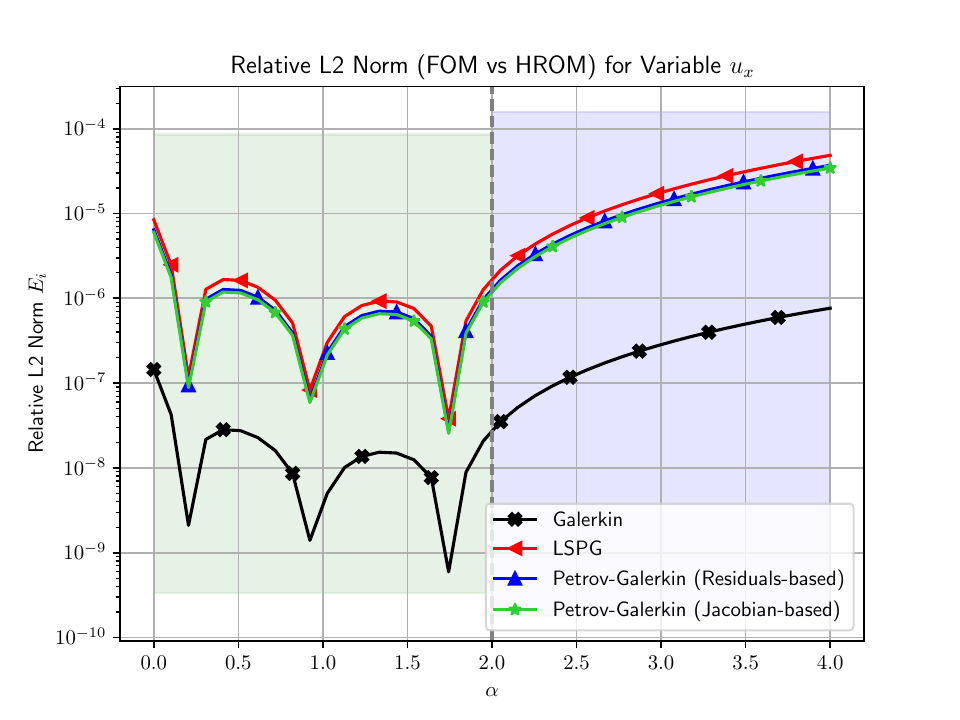}}
        \caption{Displacement in x}
        \label{fig:RelativeL2Norm_FOMvsHROM_DISPLACEMENT_X}
    \end{subfigure}%
    \begin{subfigure}{.5\textwidth}
        \centering
        \fbox{\includegraphics[width=0.9\linewidth]{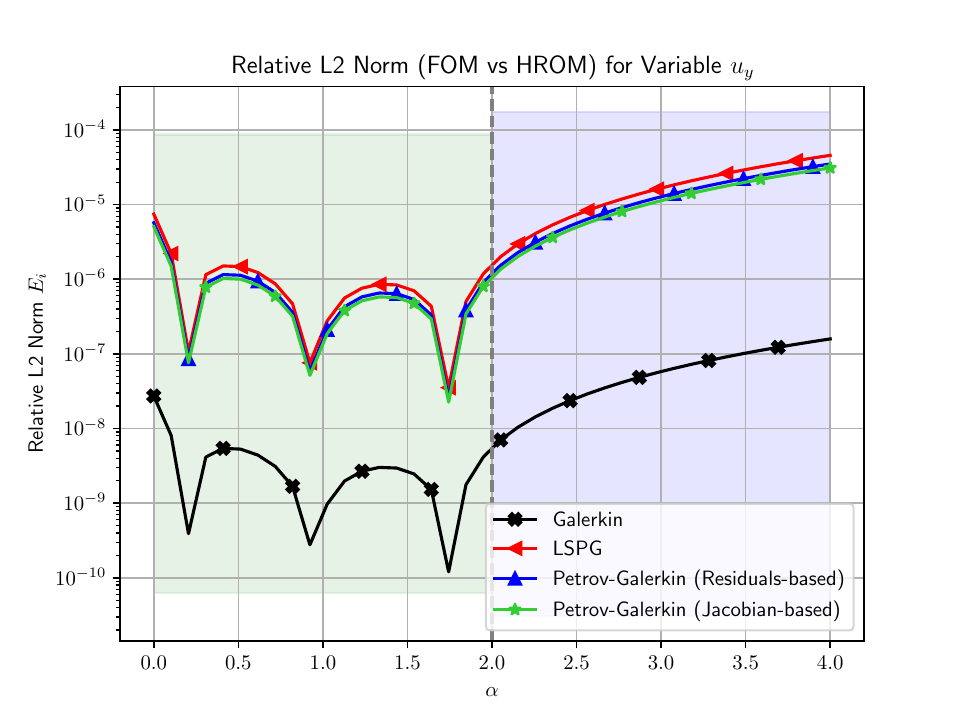}}
        \caption{Displacement in y}
        \label{fig:RelativeL2Norm_FOMvsHROM_DISPLACEMENT_Y}
    \end{subfigure}
    \caption{Relative L2 Norm between the FOM and the HROM against the scaling parameter $\alpha$. The vertical line at $\alpha$ = 2.0 denotes the boundary between the training ($\alpha$ $\in$ [0.0, 2.0], shaded green) and testing phases ($\alpha$ $\in$ (2.0, 4.0], shaded blue). Each line represents a different strategy used in the model reduction process. The subfigures represent two distinct variables of interest: (a) Displacement in x, and (b) Displacement in y.}
    \label{fig:RelativeL2Norm_FOMvsHROM_DISPLACEMENT}
\end{figure}

\subsection{Convection-diffusion Case}
In this study, we solved a 2D convection-diffusion problem on a non-uniform mesh comprising 23204 elements and 11603 nodes (degrees of freedom). This problem, characterized by a transient rotating pulse, is adapted from \cite{Huerta2003}. The transient convection-diffusion equation is expressed as:
\begin{equation}
\frac{\partial u}{\partial t} + \mathbf{a} \cdot \nabla u - \nabla \cdot (\varepsilon \nabla u) = s \quad \text{in } \Omega = [0, 1] \times [0, 1].
\end{equation}
\noindent In this equation, $u$ represents the unknown scalar field, $\frac{\partial u}{\partial t}$ is the rate of change of this field over time, and $\mathbf{a} \cdot \nabla u$ is the convective term where $\mathbf{a}=(-y,x)$ is the velocity field. The term $\nabla \cdot (\varepsilon \nabla u)$ represents diffusion, and $s$ is the source term defined as:

\begin{equation}
s = \left\{
\begin{array}{ll}
10 \times \exp\left(-50\left(x^2+y^2-\frac{1}{2}\right)^2\right) & \text{if } \sqrt{x^2 + y^2} < 1, \\
0 & \text{otherwise.}
\end{array}
\right.
\end{equation}

\noindent The boundary conditions are set as $u = 0$ on $\partial \Omega$ (the boundary of $\Omega$), and $u = 0$ at $t = 0$.\\
This problem is chosen due to its convection-dominant nature. For further details on the problem setup, please refer to \cite{Huerta2003}.\\ \\
The reduced order model was trained using properties of two different materials: Ethylene Glycol and SAE 30 Engine Oil. Ethylene Glycol has a density of approximately 1110 kg/m³, a thermal conductivity of 0.253 W/(m*K), and a specific heat of 2412 J/(kg*K). SAE 30 Engine Oil, on the other hand, has a density of approximately 875 kg/m³, a thermal conductivity of 0.15 W/(m*K), and a specific heat of 2092 J/(kg*K).\\
In the testing phase, the reduced order model was extrapolated to a third material, Glycerol. Glycerol has a density of approximately 1260 kg/m³, a thermal conductivity of 0.286 W/(m*K), and a specific heat of 2430 J/(kg*K). The primary goal of this experiment is to assess the ability of the reduced order model to generalize over varying material parameters. The time for the convection-diffusion problem was defined from 0 to 5 seconds.
\paragraph{Reduction}
The reduction process for the convection-diffusion problem was carried out with a tolerance level of $\epsilon_{u}=10^{-3}$, which was used to determine the dimensionality of the latent variables $\mathbf{\hat{u}}$. Under this tolerance, the number of modes in the right reduced order basis (ROB) $\boldsymbol{\Phi}$ was 18. The Petrov-Galerkin methods, with singular value decomposition (SVD) tolerances of $\epsilon_{\mathbf{R}}=\epsilon_{\mathbf{J}}= 10^{-3}$, resulted in a left ROB, $\boldsymbol{\Psi}$, comprising 18 modes for the Jacobian-based approach and 19 modes for the Residual-based approach. The empirical cubature method (ECM), which was utilized in all the projection strategies, was configured with a tolerance of $10^{-6}$.
\paragraph{Discussion}
Our model's initial validation utilized training materials, specifically Ethylene Glycol and SAE 30 Engine Oil. The resultant mean relative L2 errors across different model reduction strategies are encapsulated in the table \ref{table:MeanRelativeL2NormsCDTrain}.

\begin{table}[h!]
\centering
\caption{Mean Relative L2 Norms for the Training Phase}
\begin{tabular}{|c|c|c|c|}
\hline
Strategy & FOM vs ROM & ROM vs HROM & FOM vs HROM \\
\hline
Galerkin & $2.323 \times 10^{-3}$ & $1.35 \times 10^{-6}$ & $1.089 \times 10^{-3}$ \\
LSPG & $2.326 \times 10^{-3}$ & $1.399 \times 10^{-5}$ & $1.640 \times 10^{-3}$ \\
Petrov-Galerkin Res. & $2.180 \times 10^{-3}$ & $1.635 \times 10^{-4}$ & $1.267 \times 10^{-3}$ \\
Petrov-Galerkin Jac. & $2.326 \times 10^{-3}$ & $1.335 \times 10^{-6}$ & $1.230 \times 10^{-3}$ \\
\hline
\end{tabular}
\label{table:MeanRelativeL2NormsCDTrain}
\end{table}

\begin{figure}[h!]
    \centering
    \begin{subfigure}{.5\textwidth}
        \centering
        \includegraphics[width=0.9\linewidth]{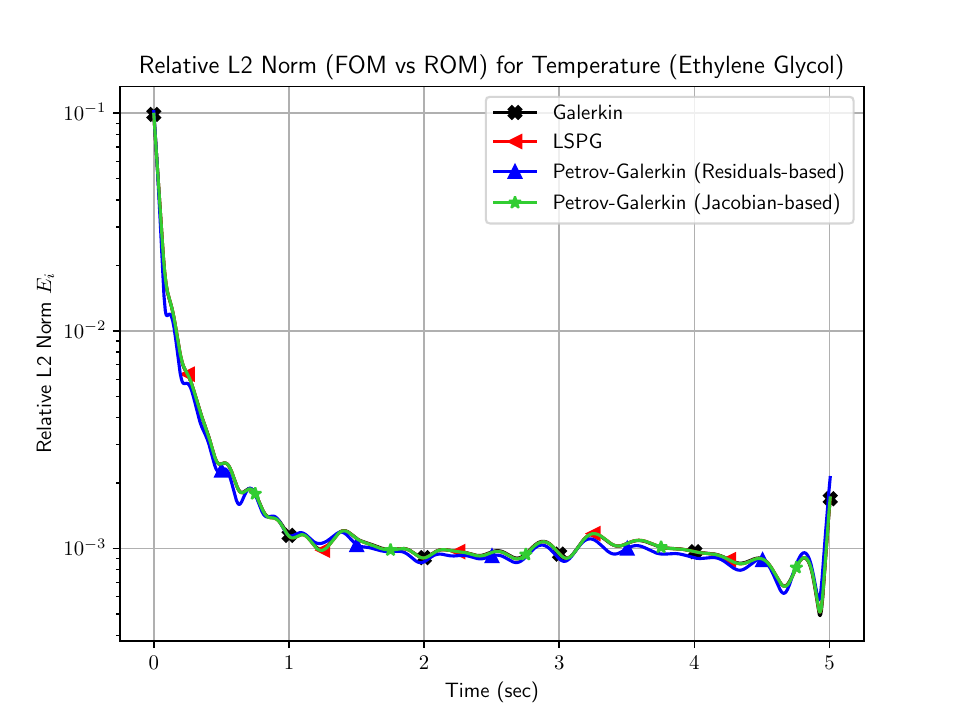}
        \caption{Ethylene Glycol}
        \label{fig:RelativeL2Norm_FOMvsROM_TEMPERATURE_EthyleneGlycol}
    \end{subfigure}%
    \begin{subfigure}{.5\textwidth}
        \centering
        \includegraphics[width=0.9\linewidth]{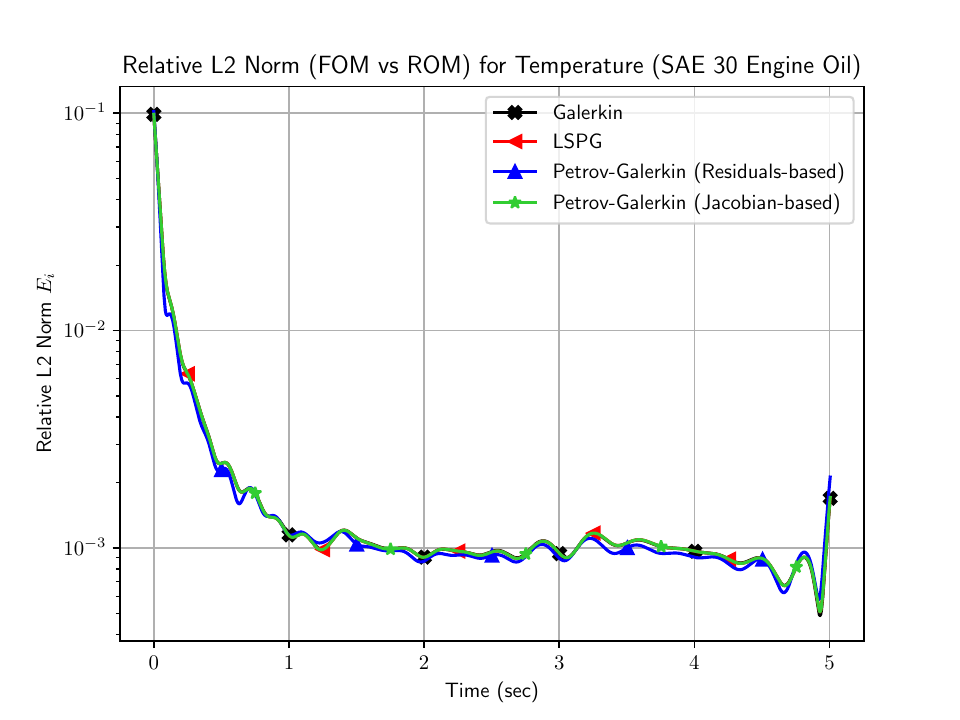}
        \caption{SAE 30 Engine Oil}
        \label{fig:RelativeL2Norm_FOMvsROM_TEMPERATURE_SAE30EngineOil}
    \end{subfigure}
    \caption{Relative L2 Norm between the FOM and ROM for different training materials. Each line represents a different strategy used in the model reduction process. The subfigures represent two distinct materials: (a) Ethylene Glycol, and (b) SAE 30 Engine Oil.}
    \label{fig:RelativeL2Norm_FOMvsROM_TEMPERATURE}
\end{figure}

\begin{figure}[h!]
    \centering
    \begin{subfigure}{.5\textwidth}
        \centering
        \includegraphics[width=0.9\linewidth]{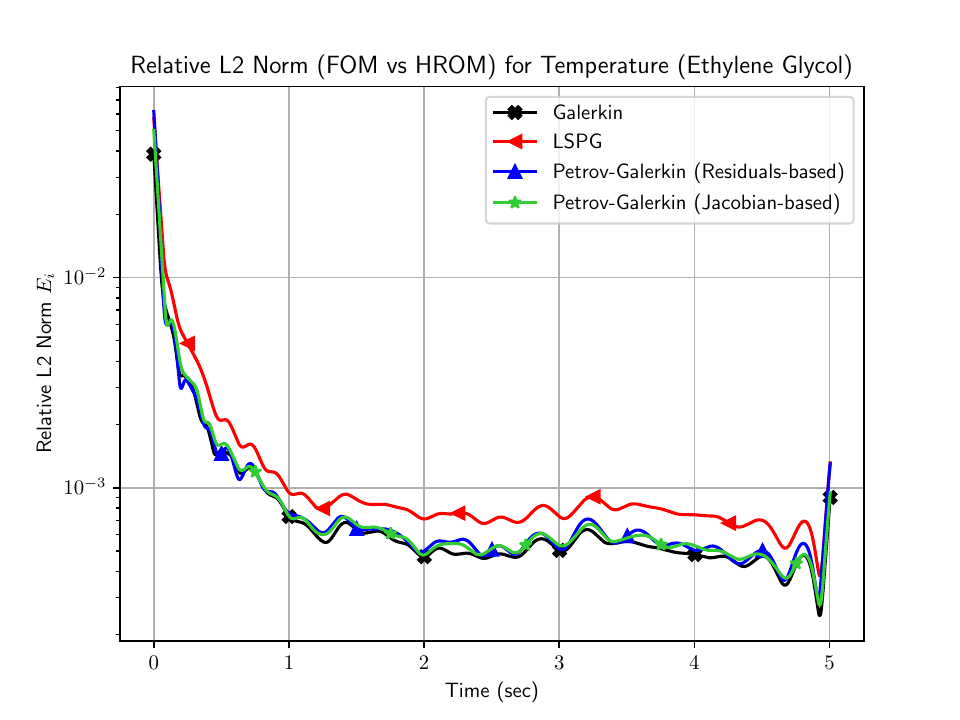}
        \caption{Ethylene Glycol}
        \label{fig:RelativeL2Norm_FOMvsHROM_TEMPERATURE_EthyleneGlycol}
    \end{subfigure}%
    \begin{subfigure}{.5\textwidth}
        \centering
        \includegraphics[width=0.9\linewidth]{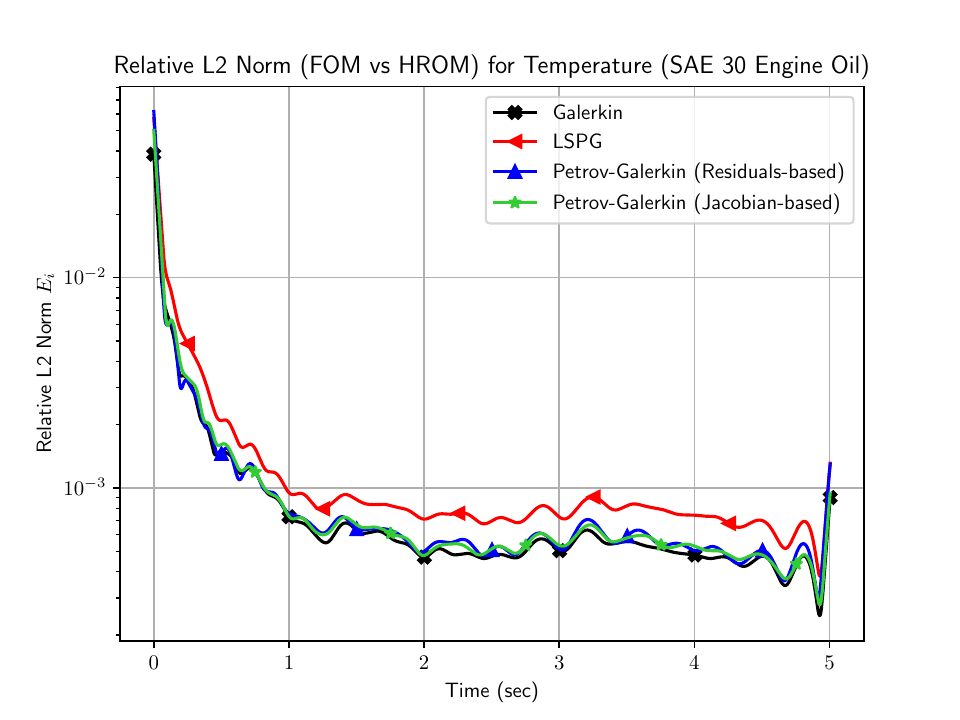}
        \caption{SAE 30 Engine Oil}
        \label{fig:RelativeL2Norm_FOMvsHROM_TEMPERATURE_SAE30EngineOil}
    \end{subfigure}
    \caption{Relative L2 Norm between the FOM and the HROM for different training materials. Each line represents a different strategy used in the model reduction process. The subfigures represent two distinct materials: (a) Ethylene Glycol, and (b) SAE 30 Engine Oil.}
    \label{fig:RelativeL2Norm_FOMvsHROM_TEMPERATURE}
\end{figure}

\noindent In contrast to the structural mechanics case, the Galerkin method, impacted by the non-SPD nature of the problem, fails to yield a clear minimum solution. Nonetheless, an interesting observation from the results is the convergence of Petrov-Galerkin strategies and the LSPG method to similar error magnitudes. Illustrating the evolution of the mean relative L2 error for varying reduction strategies concerning the temperature variable, Figures \ref{fig:RelativeL2Norm_FOMvsROM_TEMPERATURE} and \ref{fig:RelativeL2Norm_FOMvsHROM_TEMPERATURE} further highlight the implications of the problem not being SPD.

\noindent Remarkably, the Petrov-Galerkin strategies not only match the LSPG's error values but also outperform them within an HROM context further presented. The effectiveness and aptness of the Petrov-Galerkin strategies in handling such complicated problems are thereby accentuated.

\begin{figure}[h]
    \centering
    \fbox{\includegraphics[width=0.8\textwidth]{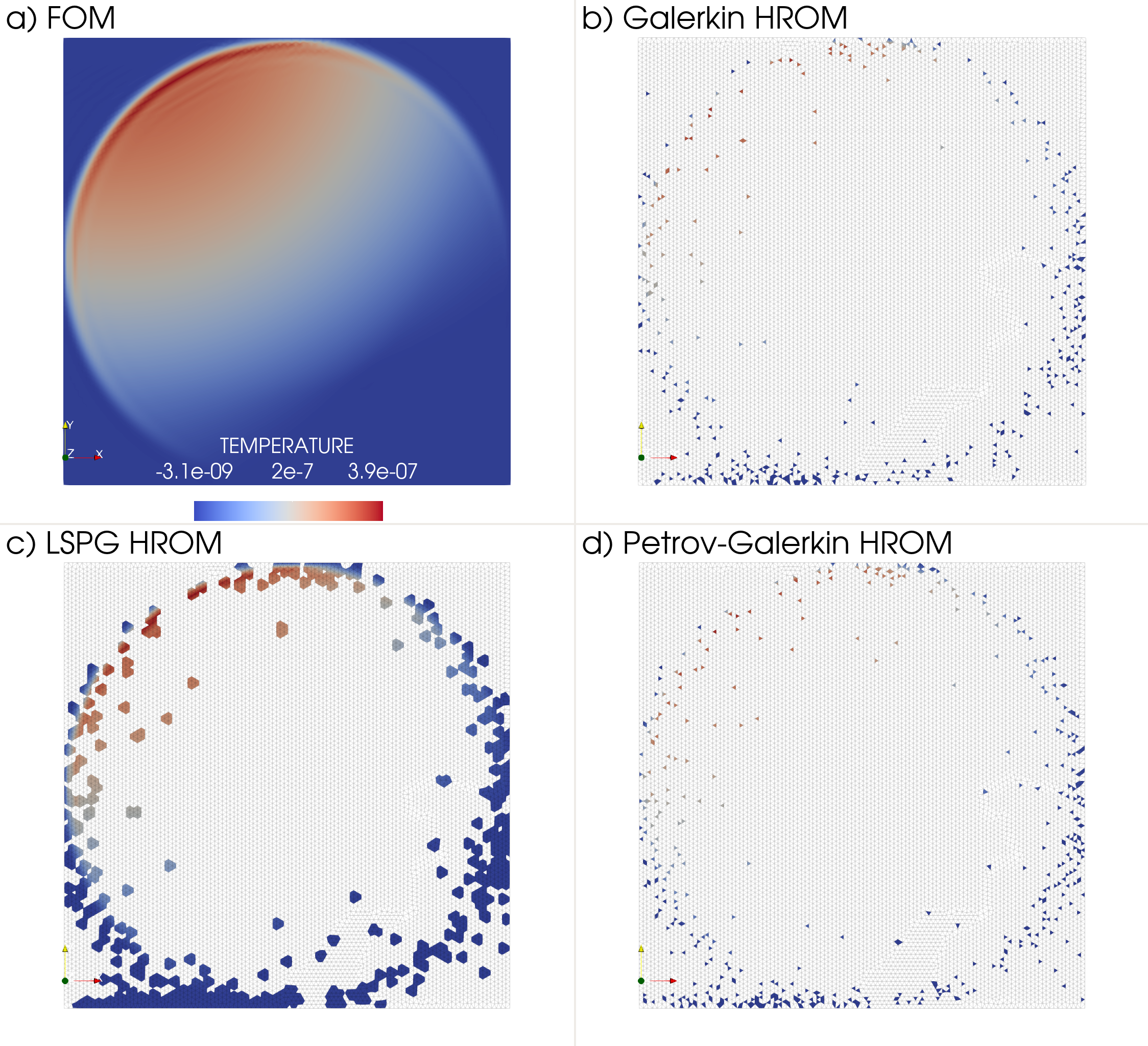}}
    \caption{Comparison of HROM solutions for different strategies in the context of the transient rotating pulse (Ethylene Glycol): (a) FOM, (b) Galerkin HROM, (c) LSPG HROM, (d) Petrov-Galerkin HROM. The figure also exhibits the different HROM meshes used for each strategy.}
    \label{fig:cd_hrom_strategies}
\end{figure}

\noindent Delving into the HROM selection process, it yielded the following element distribution: the Galerkin strategy incorporated 365 elements (representing 1.57\% of the total 23204 elements), the LSPG strategy included 358 elements (1.54\% of the total), the Petrov-Galerkin Residual-based strategy integrated 462 elements (1.99\% of the total), and the Petrov-Galerkin Jacobian-based strategy adopted 439 elements (1.89\% of the total). It is noteworthy that, despite seemingly requiring fewer elements, the LSPG strategy depends on information from surrounding elements, necessitating a complementary mesh. The LSPG strategy further results in a marked increase in the effective number of elements, growing from 358 to 3265, approximately 9.1 times the original selection. This inflated selection constitutes 14.07\% of the original FOM with 23204 elements, a significant increase from the initial 1.54\%, as depicted in Fig.\ref{fig:cd_hrom_strategies}.

\noindent In contrast, the Petrov-Galerkin strategies hold an advantage as they prevent this rise in elements while still providing a solution with a minimum residual comparable to the LSPG strategy.

\noindent Having delved into the complexities of element distribution within different strategies, let's now shift our focus to another critical metric: the total speedups achieved for the ROMs and HROMs across these strategies. This will provide us with another important perspective on the comparative performance of these methods. Table \ref{tab:speedup_and_elements} below encapsulates the total average speedups for the ROM and HROM using different strategies.

\begin{table}[h!]
\begin{center}
\begin{tabular}{|c|c|}
\hline
Strategy & Total Speedup (average over materials) \\
\hline
ROM Galerkin & 4.04 \\
HROM Galerkin & 49.0 \\
ROM LSPG & 4.33  \\
\cellcolor{gray!25}HROM LSPG & \cellcolor{gray!25}21.35 \\
ROM Petrov-Galerkin Res. & 3.63\\
HROM Petrov-Galerkin Res. & 35.39 \\
ROM Petrov-Galerkin Jac. & 3.64 \\
\cellcolor{gray!25}HROM Petrov-Galerkin Jac. & \cellcolor{gray!25}37.44 \\
\hline
\end{tabular}
\caption{Average speedups for ROM and HROM using different strategies, and the percentage of elements selected for HROM with respect to the FOM.}
\label{tab:speedup_and_elements}
\end{center}
\end{table}

\begin{figure}[h!]
    \centering
    \begin{subfigure}{.5\textwidth}
        \centering
        \includegraphics[width=0.9\linewidth]{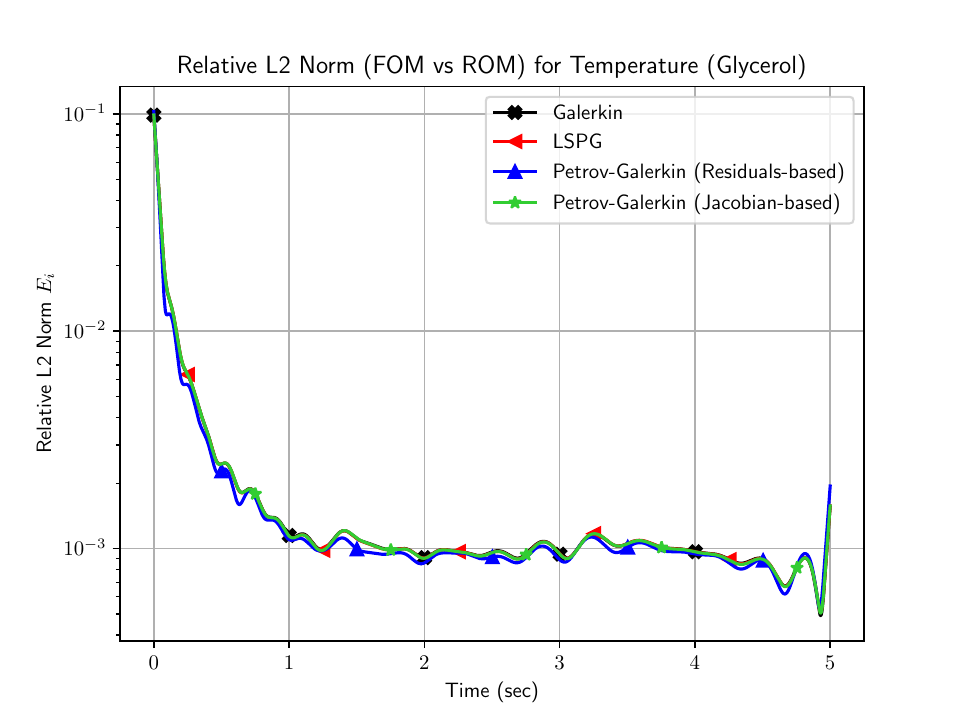}
        \caption{FOM vs ROM}
        \label{fig:RelativeL2Norm_FOMvsROM_TEMPERATURE_Glycerol_Test}
    \end{subfigure}%
    \begin{subfigure}{.5\textwidth}
        \centering
        \includegraphics[width=0.9\linewidth]{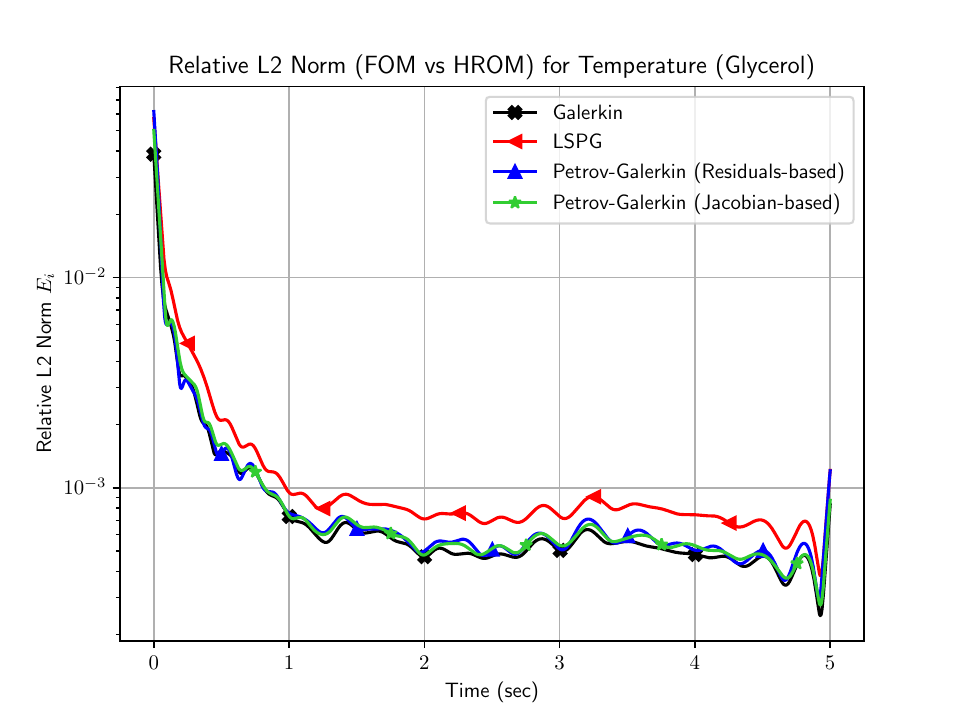}
        \caption{FOM vs HROM}
        \label{fig:RelativeL2Norm_FOMvsHROM_TEMPERATURE_Glycerol_Test}
    \end{subfigure}
    \caption{Relative L2 Norm between the FOM and ROM or HROM for the testing material Glycerol. Each line represents a different strategy used in the model reduction process.}
    \label{fig:RelativeL2Norm_TEMPERATURE_Test}
\end{figure}

\noindent After establishing our model's performance with the training materials, we expanded our analysis to encompass a material distinct from the training set, specifically, Glycerol. This rigorous test allows us to evaluate our model's ability in predicting the behavior of materials beyond its training repertoire. Delving into the comparative examination of mean relative L2 norms for different model reduction strategies applied to Glycerol, the table \ref{table:MeanRelativeL2NormsCDTest} below encapsulates these findings. It lays out the mean relative L2 norms for the temperature variable.
\begin{table}[h!]
\centering
\caption{Mean Relative L2 Norms for the Testing Phase with Glycerol}
\begin{tabular}{|c|c|c|c|}
\hline
Strategy & FOM vs ROM & ROM vs HROM & FOM vs HROM \\
\hline
Galerkin & $2.324 \times 10^{-3}$ & $1.15 \times 10^{-6}$ & $1.090 \times 10^{-3}$ \\
LSPG & $2.327 \times 10^{-3}$ & $1.41 \times 10^{-5}$ & $1.640 \times 10^{-3}$ \\
Petrov-Galerkin Res. & $2.167 \times 10^{-3}$ & $1.22 \times 10^{-14}$ & $1.267 \times 10^{-3}$ \\
Petrov-Galerkin Jac. & $2.326 \times 10^{-3}$ & $6.07 \times 10^{-15}$ & $1.231 \times 10^{-3}$ \\
\hline
\end{tabular}
\label{table:MeanRelativeL2NormsCDTest}
\end{table}
Our exploration of Glycerol test results underlines the robust performance of our reduced order model across diverse materials with comparable properties. Figures \ref{fig:RelativeL2Norm_FOMvsROM_TEMPERATURE_Glycerol_Test} and \ref{fig:RelativeL2Norm_FOMvsHROM_TEMPERATURE_Glycerol_Test} visualize the trajectory of the mean relative L2 error for various reduction strategies in relation to the temperature variable of the test material, Glycerol.

\noindent While the Galerkin strategy for HROM marks high speedup in this non-SPD problem and provides a benchmark, we must heed the fact that the Galerkin method does not guarantee convergence to the minimum solution for all non-SPD problems. Consequently, it could potentially induce instabilities \cite{Carlberg2017}. Despite its commendable performance in the given context, its reliability may falter in other scenarios. Our prime interest lies in comparing and contrasting the LSPG strategy with the Petrov-Galerkin strategies. At first glance, the LSPG strategy may appear more resourceful due to its more economical selection of elements from the FOM. However, its need for a larger effective selection driven by the necessary complementary mesh eventually cuts into its computational speedup.

\noindent Conversely, the Petrov-Galerkin strategies deliver substantial speedups while harnessing a relatively lean fraction of FOM elements. Impressively, these strategies achieve a solution fidelity comparable to the LSPG method but boast superior computational efficiency. This performance delineates a desirable balance between computational speedup and approximation accuracy, thereby hinting at the potential edge of Petrov-Galerkin strategies in non-SPD problem settings.

\subsection{Fluid Dynamics Case}
The 2D CFD simulation for the incompressible Navier-Stokes equations was performed to solve the well-known flow past a cylinder benchmark for specific base velocities. The problem geometry consists of a 5 x 2 m channel with a non-slip cylinder of 0.2 m diameter located at coordinates (1.25,0.5). The top and bottom walls are also non-slip, and the pressure is fixed along the right edge. The time-dependent parabolic inlet function is applied at the left edge, with a sinusoidal ramp-up applied to the inlet function from 0.0 to 1.0 s. This inlet function is defined as:

\begin{equation*}
v(t)=
\begin{cases}
v_{b}y(1-y) sin(\frac{\pi t}{2}), & t\in[0,1) \\
v_{b}y(1-y), & t\in[1,T]
\end{cases}
\end{equation*}

\noindent For the monitoring of the results, a probe is positioned downstream from the cylinder’s trailing edge at coordinates (1.43, 0.52). This probe serves to gather key data from the flow dynamics.\\ \\

\noindent The simulations are run with base velocities ($v_{b}$) of $4.0 \frac{m}{s}$ and $6.0 \frac{m}{s}$ for a total of 30 seconds for the training phase, and with a base velocity of $5.0 \frac{m}{s}$ for 40 seconds during the testing phase. These base velocities correspond to different Reynolds numbers, calculated using the formula:

\begin{equation*}
Re = \frac{v_{avg} * \rho * D}{\eta},
\end{equation*}

\noindent where $\eta = 0.002 \frac{kg}{m\cdot s}$ is the dynamic viscosity, $\rho = 1 \frac{kg}{m^3}$ is the fluid density, $D=0.2$ m is the diameter of the cylinder, and $v_{avg}$ is the average velocity, calculated as $\frac{v_{b}}{6}$.\ \
The calculated Reynolds numbers for each base velocity are presented in Table \ref{ReTable}.

\begin{table}[h!]
\centering
\begin{tabular}{|c|c|}
\hline
\textbf{$v_{b}$ ($\frac{m}{s}$)} & \textbf{Re} \\
\hline
4.0 & 66.7 \\
\cellcolor{gray!25}5.0 & \cellcolor{gray!25} 83.3 \\
6.0 & 100 \\
\hline
\end{tabular}
\caption{The calculated Reynolds numbers for each base velocity. The test scenario ($v_{b} = 5.0 \frac{m}{s}$) is shaded in grey.}
\label{ReTable}
\end{table}

\begin{figure}[h!]
\centering
\fbox{\includegraphics[width=0.7\linewidth]{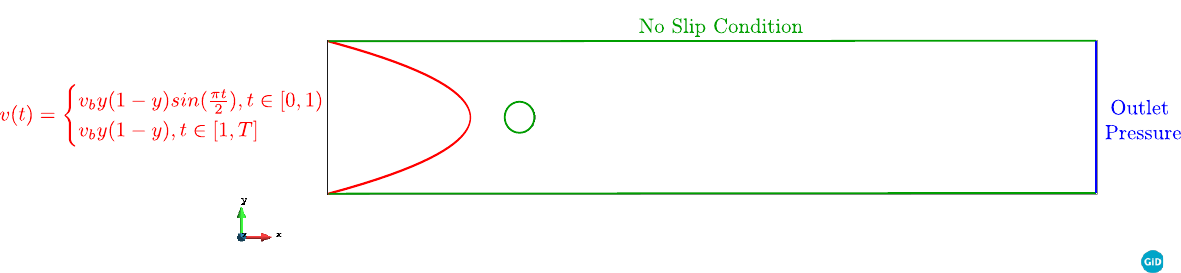}}
\caption{Flow past a cylinder 2D.}
\label{fig:Flow past a cylinder 2D}
\end{figure}

\noindent The time step is 0.1 second, and the mesh consists of  68298 linear triangular elements with 34149 nodes (102447 degrees of freedom). Figure \ref{fig:Flow past a cylinder 2D} illustrates the problem setup and boundary conditions. By simulating this benchmark, we aim to evaluate the ability of our reduced-order models to reconstruct the flow behavior at different Reynolds numbers, derived from various base velocities.

\begin{figure}[h!]
\centering
\begin{subfigure}{.5\textwidth}
\centering
\includegraphics[width=0.9\linewidth]{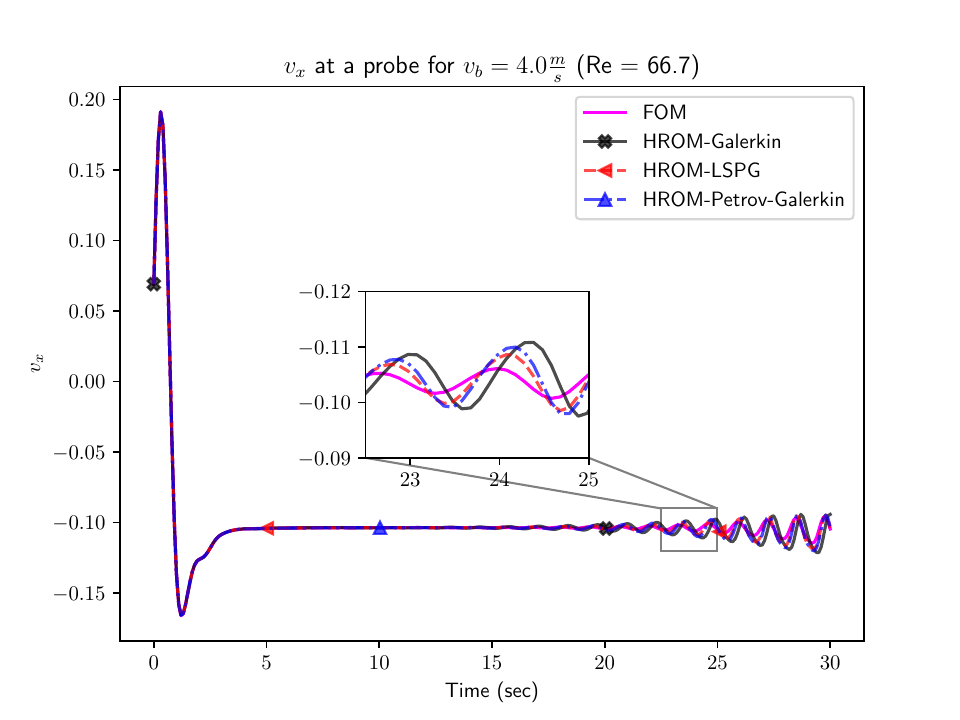}
\caption{Velocity component in the x-direction at a selected probe node for the FOM and various HROM strategies ($v_b = 4.0\frac{m}{s}$)}
\label{fig:VELOCITY_X_Node_9599_mu_4.0}
\end{subfigure}%
\begin{subfigure}{.5\textwidth}
\centering
\includegraphics[width=0.9\linewidth]{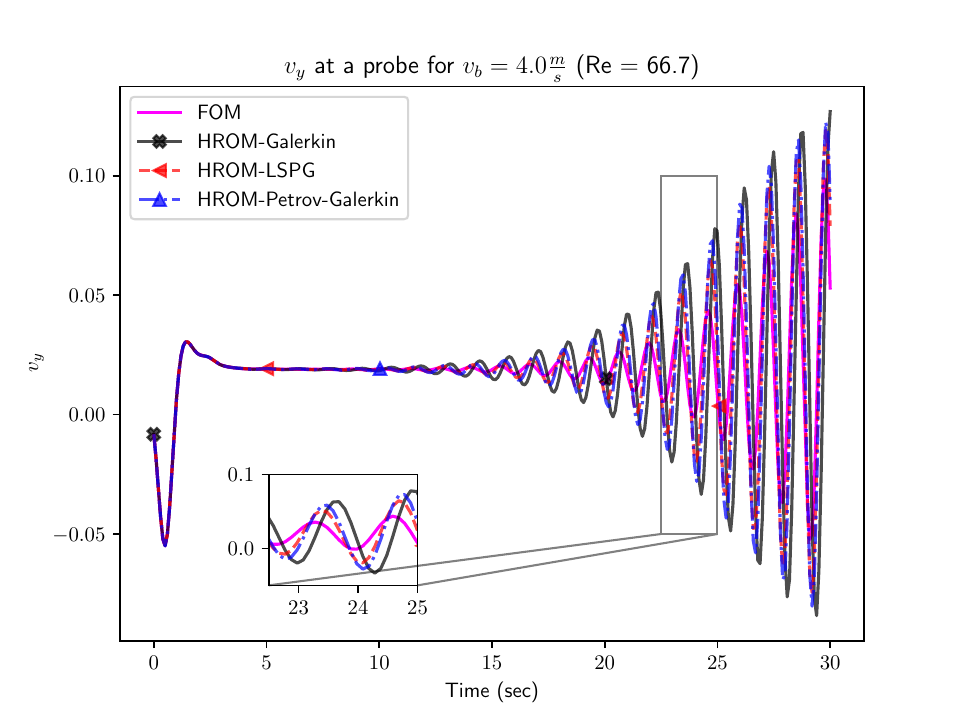}
\caption{Velocity component in the y-direction at a selected probe node for the FOM and various HROM strategies ($v_b = 4.0\frac{m}{s}$)}
\label{fig:VELOCITY_Y_Node_9599_mu_4.0}
\end{subfigure}
\caption{Velocity profiles at a selected probe node for the FOM and HROM using Galerkin, LSPG, and Petrov-Galerkin strategies. The velocities correspond to a Reynolds number of 66.7 ($v_{b} = 4.0 \frac{m}{s}$) and represent a trainning parameter.. Zoomed-in regions are provided to visualize specific details.}
\label{fig:Velocity_Profiles_Node_Probe}
\end{figure}

\begin{figure}[h!]
\centering
\begin{subfigure}{.5\textwidth}
\centering
\includegraphics[width=0.9\linewidth]{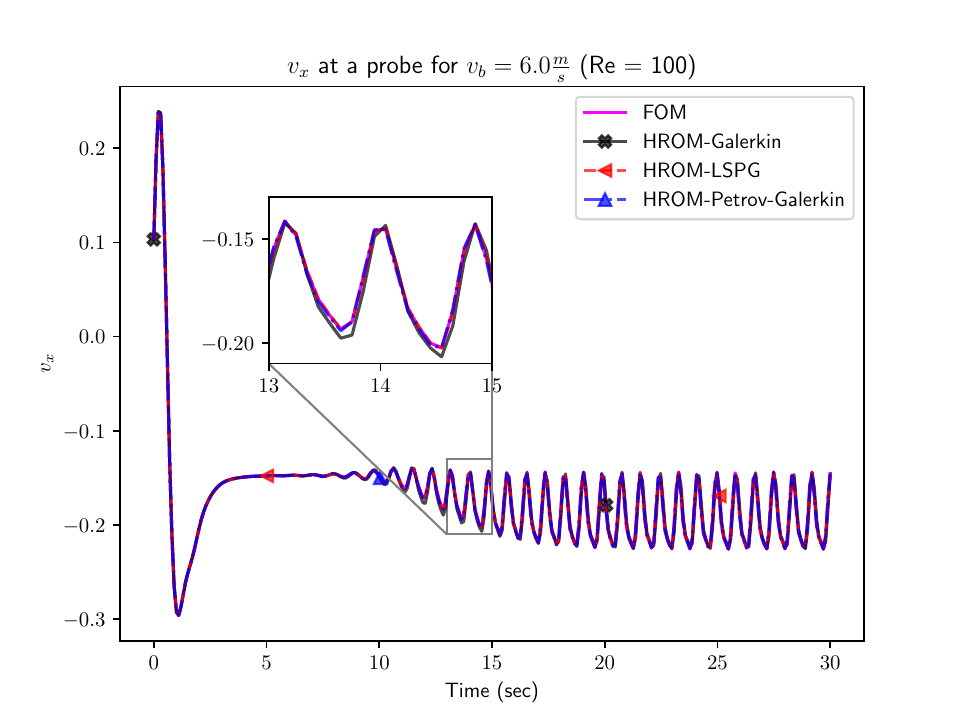}
\caption{Velocity component in the x-direction at a selected probe node for the FOM and various HROM strategies ($v_b = 6.0\frac{m}{s}$)}
\label{fig:VELOCITY_X_Node_9599_mu_6.0}
\end{subfigure}%
\begin{subfigure}{.5\textwidth}
\centering
\includegraphics[width=0.9\linewidth]{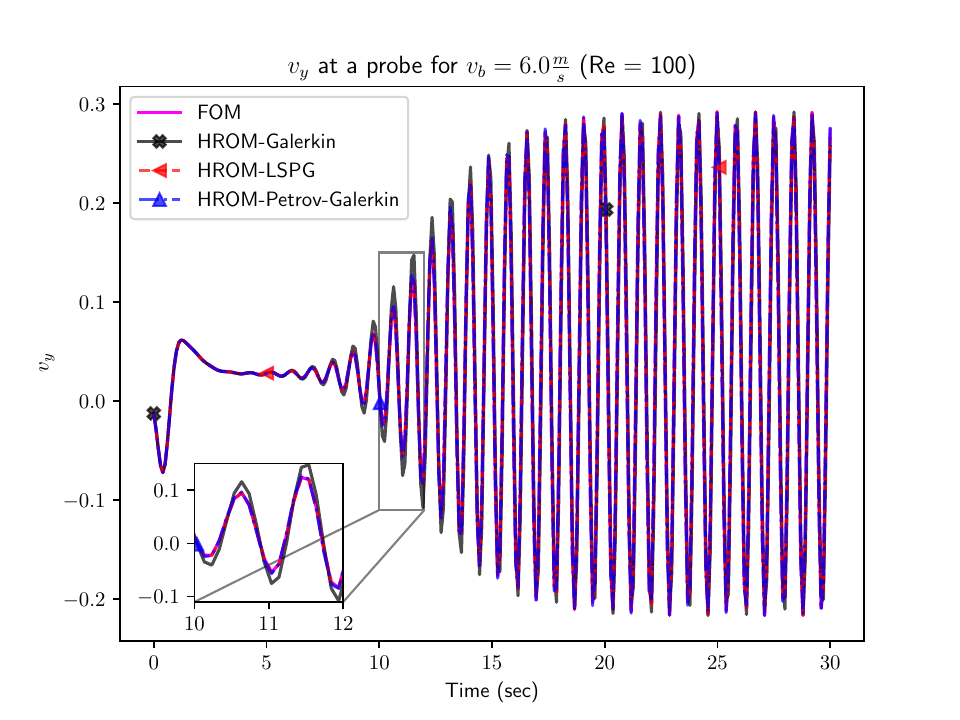}
\caption{Velocity component in the y-direction at a selected probe node for the FOM and various HROM strategies ($v_b = 6.0\frac{m}{s}$)}
\label{fig:VELOCITY_Y_Node_9599_mu_6.0}
\end{subfigure}
\caption{Velocity profiles at a selected probe node for the FOM and HROM using Galerkin, LSPG, and Petrov-Galerkin strategies. The velocities correspond to a Reynolds number of 100 ($v_{b} = 6.0 \frac{m}{s}$) and represent a trainning parameter.. Zoomed-in regions are provided to visualize specific details.}
\label{fig:Velocity_Profiles_Node_Probe_6.0}
\end{figure}

\paragraph{Reduction}
In this specific analysis, tolerance levels were carefully selected to obtain a sufficiently good approximation of the fluid-dynamics problem at hand, to avoid introducing high phase and amplitude errors. A tolerance level of $\epsilon_u = 10^{-4}$ was adopted to ascertain the dimensionality of the latent variables $\mathbf{\hat{u}}$. This resulted in the construction of a right ROB $\boldsymbol{\Phi}$ that incorporated 98 modes. For the Petrov-Galerkin approach, particularly the Residual-based method, a SVD tolerance of $\epsilon_{\mathbf{R}} = 5\times 10^{-4}$ was employed to prevent the left ROB from becoming excessively large, and yielded a left ROB $\boldsymbol{\Psi}$ consisting of 153 modes. For the left ROB, only the Residual-based approach was employed in this case to minimize the memory load during the second training phase. This decision was strategic, as the Jacobian-based method would have resulted in a snapshots matrix for the left ROB growing linearly with the number of modes in the right ROB. In contrast, the Residual-based approach's snapshots matrix expands linearly with the number of nonlinear iterations, which are significantly fewer in this case than the number of modes. This strategy facilitated a more efficient training process without sacrificing the accuracy of the results. The empirical cubature method (ECM), deployed across all projection strategies, was configured with a tolerance of $10^{-5}$.
\paragraph{Discussion}
With the objective of appraising the resilience and accuracy of our training model, we embarked on a validation study, employing the training parameters $v_{b}$ of $4.0 \frac{m}{s}$ ($Re = 66.7$) and $6.0 \frac{m}{s}$ ($Re = 100$). These specific values were judiciously selected to encompass a varied, yet illustrative range of the prospective parameter space wherein the model would be anticipated to perform.\\ \\
\noindent In this context, it is important to highlight certain aspects of the outcome that stem directly from the unique properties of the boundary conditions (e.g. the inlet ramp-up) and parametric settings for this case. Given the complexity of the problem, the model didn't showcase a substantial reduction in the information required for an accurate representation. Specifically, out of 600 snapshots collected for the training set, the optimal right ROB required 98 modes to accurately represent both the phase and amplitude of the dynamics, devoid of high spikes and significant errors. This translates into the need for roughly one-sixth of the total information of the problem. While such a requirement may lead to potential limitations in the performance of both the ROM and the HROM, it is pivotal to note that these outcomes do not undermine the value of our research. Rather, they provide an opportunity to gain insights into the nature of complex systems.\\ \\
Moreover, the primary intention of this paper is not focused on achieving the highest efficiency of information reduction, but rather on demonstrating and accentuating the performance of the proposed Petrov-Galerkin HROM in comparison to the LSPG approach for a general case. Even under less-than-optimal circumstances, our findings shed valuable light on the versatility and potential of the proposed HROM framework. One important aspect to highlight is the relative performance of the different reduced order models. Throughout the simulation, it was observed that both the LSPG and Petrov-Galerkin cases consistently presented results more closely aligned with those of the FOM, compared to the Galerkin ROM. This outcome underscores the superior performance of the Petrov-Galerkin and LSPG models in replicating the detailed fluid dynamics captured by the FOM, even under vortical flow conditions.\\ \\
Furthermore, it is noteworthy that as the size of the left ROB is increased for the Petrov-Galerkin model, its outcomes should progressively align with those of the LSPG model. In other words, with a sufficiently large left ROB, the Petrov-Galerkin model would essentially replicate the LSPG results, underlining the inherent congruence between these two ROM methodologies when deployed with adequate basis sets. This characteristic reinforces their capability for delivering reliable and precise simulations, consistent with the FOM, for a wide range of fluid dynamics problems. The outcomes of these comparisons are visually represented in Figures \ref{fig:VELOCITY_X_Node_9599_mu_4.0}, \ref{fig:VELOCITY_Y_Node_9599_mu_4.0}, \ref{fig:VELOCITY_X_Node_9599_mu_6.0}, and \ref{fig:VELOCITY_Y_Node_9599_mu_6.0}.\\

\begin{figure}[h]
\centering
\fbox{\includegraphics[width=1.0\textwidth]{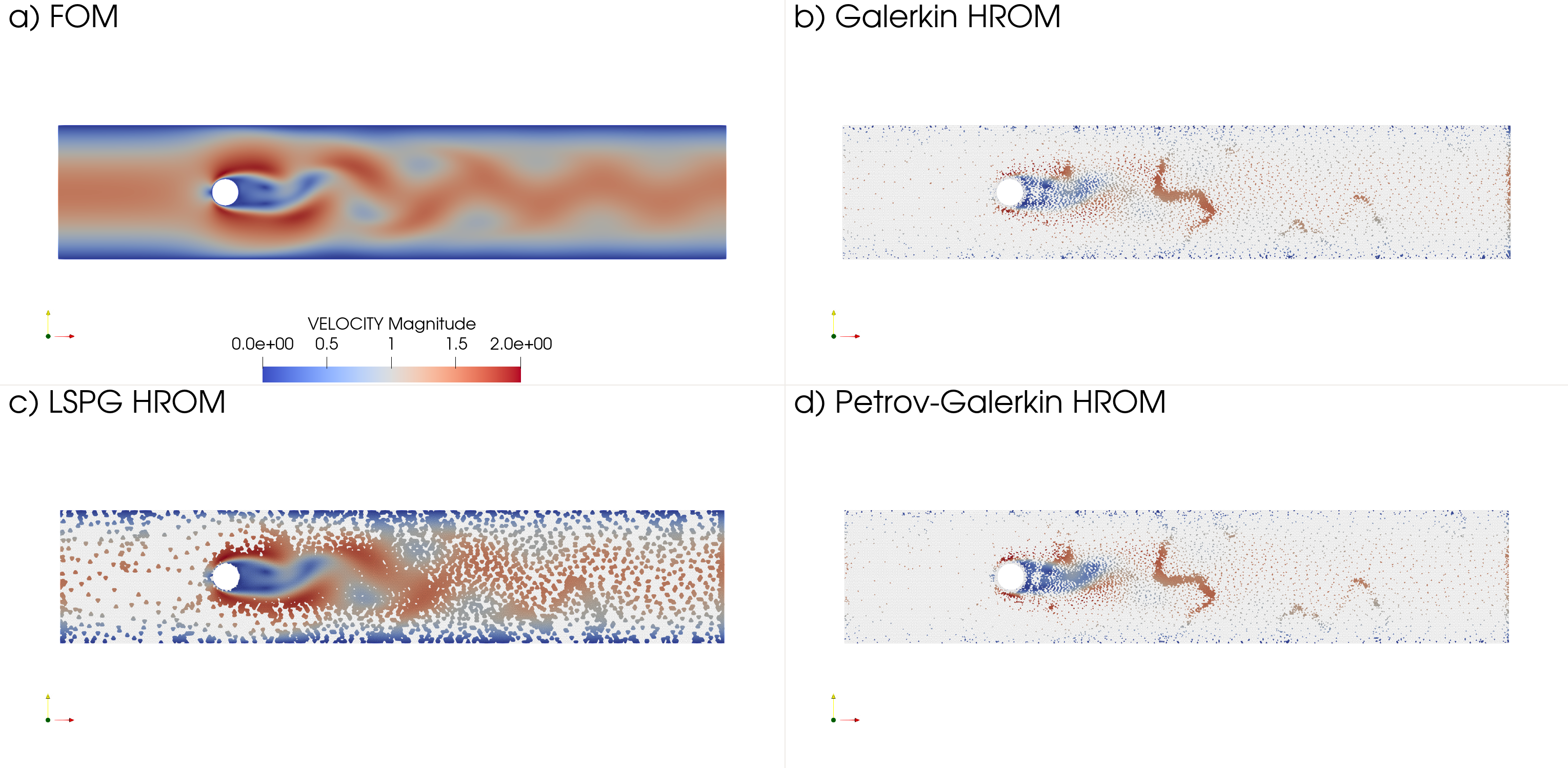}}
\caption{Comparison of HROM solutions for different strategies in the context of the transient flow past a cylinder ($v_b = 6.0 \frac{m}{s}$): (a) FOM, (b) Galerkin HROM, (c) LSPG HROM, (d) Petrov-Galerkin HROM. The figure also showcases the different HROM meshes used for each strategy.}
\label{fig:fd_hrom_strategies}
\end{figure}

\noindent The HROM selection process yielded the following element distribution: the Galerkin strategy incorporated 6216 elements (representing about 9.10\% of the total 68298 elements in the FOM), the LSPG strategy included 5914 elements (8.66\% of the total), and the Petrov-Galerkin strategy integrated 5829 elements (about 8.53\% of the total). Though the LSPG strategy seems to require fewer elements, it leverages information from surrounding elements, hence necessitating more elements in a complementary mesh. Consequently, the effective number of elements expands from 5914 to 32838, about 5.55 times the original selection, constituting 48.07\% of the original FOM. This surge represents a substantial increase from the initial 8.66\%, as depicted in Fig.\ref{fig:fd_hrom_strategies}. In contrast, the Petrov-Galerkin strategies avert this surge in elements while still providing a solution with a minimum residual comparable to the LSPG strategy.\\

\noindent With a comprehensive understanding of the element distribution across different strategies, it is equally important to delve into another significant performance metric: the total speedups achieved for the HROM across these strategies. This analysis will provide us with a valuable perspective on the comparative efficiency of these methods. Table \ref{tab:hrom_speedup} encapsulates the total speedups for the HROM using different strategies. \\

\begin{table}[h!]
\begin{center}
\begin{tabular}{|c|c|}
\hline
Strategy & Total Speedup \\
\hline
HROM Galerkin & 3.29 \\
\cellcolor{gray!25}HROM LSPG & \cellcolor{gray!25}1.81 \\
\cellcolor{gray!25}HROM Petrov-Galerkin & \cellcolor{gray!25}3.15 \\
\hline
\end{tabular}
\caption{Total speedups for HROM using different strategies.}
\label{tab:hrom_speedup}
\end{center}
\end{table}

\begin{figure}[h!]
\centering
\begin{subfigure}{.5\textwidth}
\centering
\includegraphics[width=0.9\linewidth]{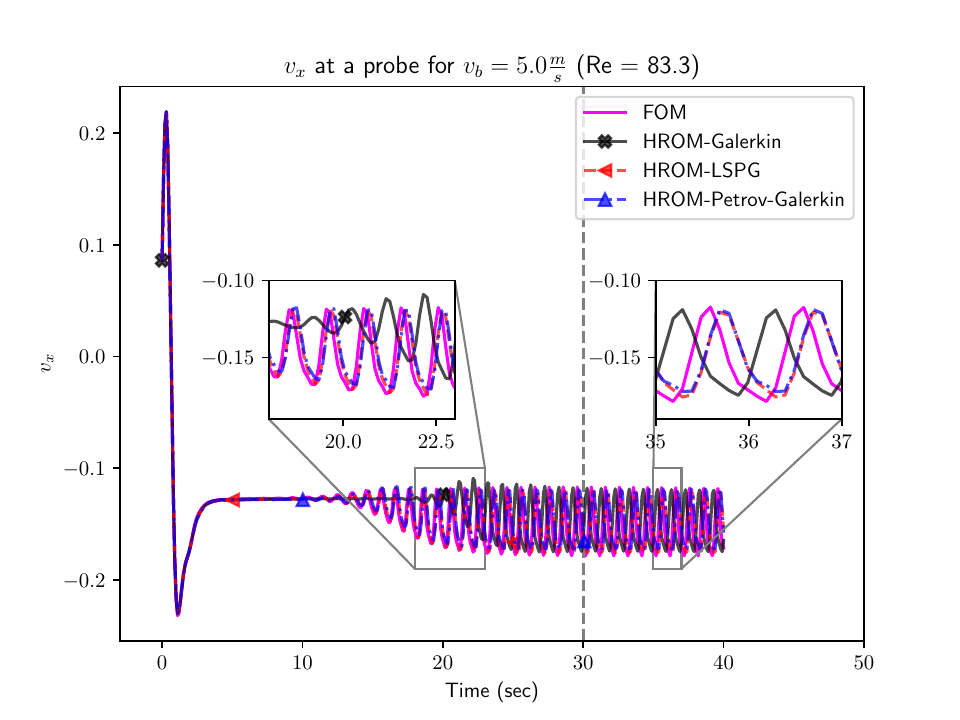}
\caption{Velocity component in the x-direction at a selected probe node for the FOM and various HROM strategies ($v_b = 5.0\frac{m}{s}$)}
\label{fig:VELOCITY_X_Node_9599_mu_5.0}
\end{subfigure}%
\begin{subfigure}{.5\textwidth}
\centering
\includegraphics[width=0.9\linewidth]{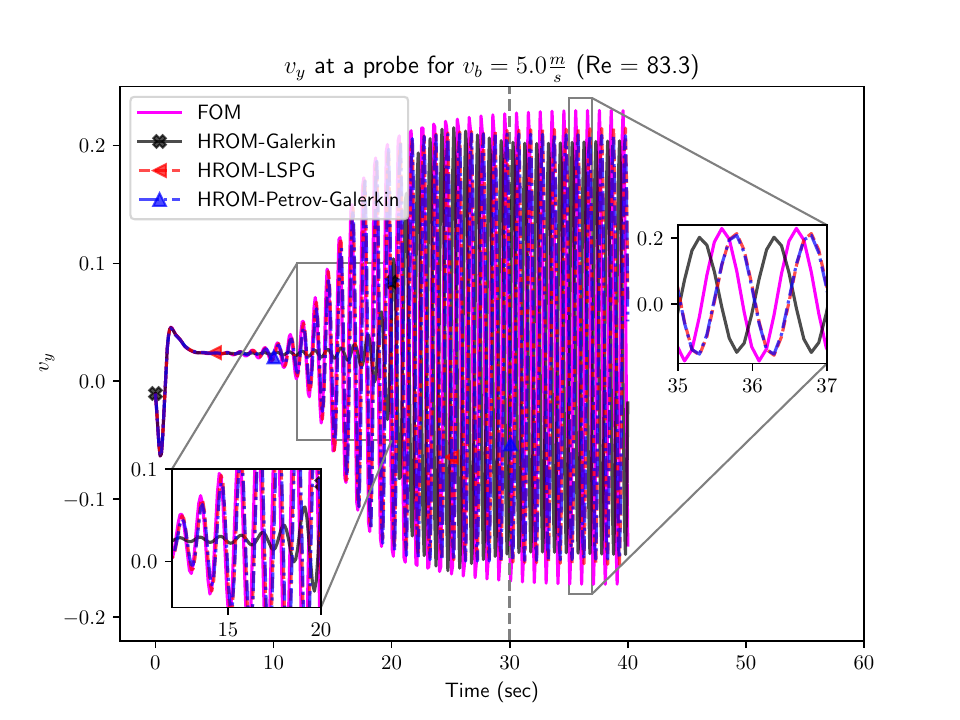}
\caption{Velocity component in the y-direction at a selected probe node for the FOM and various HROM strategies ($v_b = 5.0\frac{m}{s}$)}
\label{fig:VELOCITY_Y_Node_9599_mu_5.0}
\end{subfigure}
\caption{Velocity profiles at a selected probe node for the FOM and HROM using Galerkin, LSPG, and Petrov-Galerkin strategies. The velocities correspond to a Reynolds number of 83.3 ($v_{b} = 5.0 \frac{m}{s}$) and represent a testing parameter. Zoomed-in regions are provided to visualize specific details. The vertical dashed gray line indicates the edge of the training horizon (time = 30 sec).}
\label{fig:Velocity_Profiles_Node_Probe_5.0}
\end{figure}

\noindent The speedups depicted in Table \ref{tab:hrom_speedup} attest to the efficiency and effectiveness of these strategies within an HROM context. Notably, the Petrov-Galerkin strategy outperforms the LSPG strategy in terms of speedup, despite delivering comparable results. This improved efficiency can be attributed to Petrov-Galerkin's ability to bypass the need for a complementary mesh, offering significant advantages when selecting an optimal strategy for problems exhibiting non-SPD operators.\\ \\
To further assess the performance of our model, we conducted additional tests for a different inflow velocity of $v_b = 5.0 \frac{m}{s}$, corresponding to a Reynolds number of 83.3. This lies between the two training Reynolds numbers, allowing us to examine the capability of our model to interpolate for a Reynolds number not explicitly included in the training set. Additionally, we tested the ability of our model to extrapolate in time for another 10 seconds, reaching up to 40 seconds instead of the 30 seconds it was trained on. This evaluation is critical to determining the robustness and adaptability of our model to predict future states beyond its training horizon. The results of these tests are visually represented in Figures \ref{fig:VELOCITY_X_Node_9599_mu_5.0} and \ref{fig:VELOCITY_Y_Node_9599_mu_5.0}.\\

\noindent The figures reveal that the LSPG and Petrov-Galerkin strategies more accurately reproduce the phase and amplitude of the FOM. In contrast, the Galerkin strategy takes more time to reach the fully developed flow, shows higher amplitude errors, and experiences a noticeable phase shift. This underscores the effectiveness of the LSPG and Petrov-Galerkin strategies in accurately capturing the system dynamics, even under interpolation and extrapolation scenarios.
\section{Conclusions}

In this study, we have proposed an innovative approach that blends two key components: an invariant left ROB $\mathbf{\Psi}$ and its natural integration with the ECM hyper-reduction method. Crucially, our framework allows the left ROB to be different from the right ROB, providing enhanced flexibility in the hyper-reduction of nonlinear methods. Our model demonstrated significant improvements in accuracy, particularly for problems with non-SPD operators. Additionally, it exhibited remarkable efficiency in generating hyper-reduced models for PG-PROMs while preserving the minimum-residual optimality offered by LSPG projection. In the diverse range of problems studied, including structural mechanics (SPD), convection-diffusion (non-SPD), and fluid dynamics (non-SPD), our model showcased robust performance. A noteworthy aspect of our findings was the comparable performance of Petrov-Galerkin strategies to LSPG, despite the former exhibiting superior computational speedups. Petrov-Galerkin strategies also managed to bypass the surge in the requirement of elements, a limitation observed in the LSPG approach due to the need for a complementary mesh. Although the introduction of our proposed PG-HPROM necessitates an additional offline training stage, its efficiency in terms of the number of selected elements equals that of Galerkin HPROMs, eliminating the need for a complementary mesh. Importantly, the model's robustness was further reinforced through validation studies, demonstrating its ability to accurately interpolate and extrapolate under conditions and timelines not explicitly included in the training set. Given these results, our approach signifies broad applicability in complex industrial models across various fields, such as structural mechanics, convection-diffusion, and fluid dynamics, especially for problems with non-SPD Jacobians. This research represents a substantial stride in the field of hyper-reduction for nonlinear methods, indicating promising potential for further improvements in computational efficiency and model fidelity in the realm of reduced-order modeling. As an expansion, our methodology could further leverage advancements in local bases and neural networks, suggesting exciting avenues for future exploration and application.

\section*{Acknowledgements}
The authors acknowledge financial support from the Spanish Ministry of Economy and Competitiveness, through the ``Severo Ochoa Programme for Centres of Excellence in R\&D'' (CEX2018-000797-S)”. We acknowledge financial support from the Generalitat de Catalunya through the FI\_SDUR-2021 grant, which provided funding for the predoctoral training of Sebastian Ares de Parga Regalado. This project has received funding from the European High-Performance Computing Joint Undertaking (JU) under grant agreement Nos 955558 and 956104. The JU receives support from the European Union’s Horizon 2020 research and innovation programme and from Spain, Germany, France, Italy, Poland, Switzerland, and Norway. Spanish Ministry of Science and Innovation (MICINN) supports the project through the project PCI2021-121945. This publication is part of the R\&D project PCI2021-121944, financed by MCIN/AEI/10.13039/501100011033 and by the ``European Union NextGenerationEU/PRTR".
\bibliographystyle{ieeetr}
\bibliography{bib.bib}

\begin{thebibliography}{10}

\bibitem{Sirovich1987}
L.~Sirovich, ``Turbulence and the dynamics of coherent structures. i. coherent
  structures,'' {\em Quarterly of Applied Mathematics}, vol.~45, 1987.

\bibitem{Balachandar1998}
S.~Balachandar, ``Turbulence, coherent structures, dynamical systems and
  symmetry,'' {\em AIAA Journal}, vol.~36, 1998.

\bibitem{Antoulas2005}
A.~C. Antoulas, ``An overview of approximation methods for large-scale
  dynamical systems,'' {\em Annual Reviews in Control}, vol.~29, 2005.

\bibitem{Cuong05}
N.~N. Cuong, K.~Veroy, and A.~T. Patera, {\em Certified Real-Time Solution of
  Parametrized Partial Differential Equations}.
\newblock 2005.

\bibitem{Rowley2004}
C.~W. Rowley, T.~Colonius, and R.~M. Murray, ``Model reduction for compressible
  flows using pod and galerkin projection,'' {\em Physica D: Nonlinear
  Phenomena}, vol.~189, 2004.

\bibitem{Benner2015}
P.~Benner, S.~Gugercin, and K.~Willcox, ``A survey of projection-based model
  reduction methods for parametric dynamical systems,'' {\em SIAM Review},
  vol.~57, 2015.

\bibitem{Carlberg2011}
K.~Carlberg, C.~Bou-Mosleh, and C.~Farhat, ``Efficient non-linear model
  reduction via a least-squares petrov-galerkin projection and compressive
  tensor approximations,'' {\em International Journal for Numerical Methods in
  Engineering}, vol.~86, 2011.

\bibitem{Everson1995}
R.~Everson and L.~Sirovich, ``Karhunen–loève procedure for gappy data,''
  {\em Journal of the Optical Society of America A}, vol.~12, 1995.

\bibitem{Chaturantabut2010}
S.~Chaturantabut and D.~C. Sorensen, ``Nonlinear model reduction via discrete
  empirical interpolation,'' {\em SIAM Journal on Scientific Computing},
  vol.~32, 2010.

\bibitem{Galbally2010}
D.~Galbally, K.~Fidkowski, K.~Willcox, and O.~Ghattas, ``Non-linear model
  reduction for uncertainty quantification in large-scale inverse problems,''
  {\em International Journal for Numerical Methods in Engineering}, vol.~81,
  2010.

\bibitem{Ryckelynck2005}
D.~Ryckelynck, ``A priori hyperreduction method: An adaptive approach,'' {\em
  Journal of Computational Physics}, vol.~202, 2005.

\bibitem{Carlberg2013}
K.~Carlberg, C.~Farhat, J.~Cortial, and D.~Amsallem, ``The gnat method for
  nonlinear model reduction: Effective implementation and application to
  computational fluid dynamics and turbulent flows,'' {\em Journal of
  Computational Physics}, vol.~242, 2013.

\bibitem{Farhat2015}
C.~Farhat, T.~Chapman, and P.~Avery, ``Structure-preserving, stability, and
  accuracy properties of the energy-conserving sampling and weighting method
  for the hyper reduction of nonlinear finite element dynamic models,'' {\em
  International Journal for Numerical Methods in Engineering}, vol.~102, 2015.

\bibitem{Joaquin2016}
J.~A. Hernández, M.~A. Caicedo, and A.~Ferrer, ``Dimensional hyper-reduction
  of nonlinear finite element models via empirical cubature,'' {\em Computer
  Methods in Applied Mechanics and Engineering}, vol.~313, 2017.

\bibitem{Joaquin2020}
J.~A. Hernández, ``A multiscale method for periodic structures using domain
  decomposition and ecm-hyperreduction,'' {\em Computer Methods in Applied
  Mechanics and Engineering}, vol.~368, 2020.

\bibitem{Barrault2004}
M.~Barrault, Y.~Maday, N.~C. Nguyen, and A.~T. Patera, ``An ‘empirical
  interpolation’ method: application to efficient reduced-basis
  discretization of partial differential equations,'' {\em Comptes Rendus
  Mathematique}, vol.~339, 2004.

\bibitem{Farhat2014}
C.~Farhat, P.~Avery, T.~Chapman, and J.~Cortial, ``Dimensional reduction of
  nonlinear finite element dynamic models with finite rotations and
  energy-based mesh sampling and weighting for computational efficiency,'' {\em
  International Journal for Numerical Methods in Engineering}, vol.~98, 2014.

\bibitem{Amsallem2012}
D.~Amsallem, M.~J. Zahr, and C.~Farhat, ``Nonlinear model order reduction based
  on local reduced-order bases,'' {\em International Journal for Numerical
  Methods in Engineering}, vol.~92, 2012.

\bibitem{Carlberg2017}
K.~Carlberg, M.~Barone, and H.~Antil, ``Galerkin v. least-squares
  petrov–galerkin projection in nonlinear model reduction,'' {\em Journal of
  Computational Physics}, vol.~330, 2017.

\bibitem{Grimberg2021}
S.~Grimberg, C.~Farhat, R.~Tezaur, and C.~Bou-Mosleh, ``Mesh sampling and
  weighting for the hyperreduction of nonlinear petrov–galerkin reduced-order
  models with local reduced-order bases,'' {\em International Journal for
  Numerical Methods in Engineering}, vol.~122, 2021.

\bibitem{Blonigan2020}
P.~J. Blonigan, K.~T. Carlberg, F.~Rizzi, M.~Howard, and J.~A. Fike, ``Model
  reduction for hypersonic aerodynamics via conservative lspg projection and
  hyper-reduction,'' 2020.

\bibitem{Shimizu2021}
Y.~S. Shimizu and E.~J. Parish, ``Windowed space–time least-squares
  petrov–galerkin model order reduction for nonlinear dynamical systems,''
  {\em Computer Methods in Applied Mechanics and Engineering}, vol.~386, 2021.

\bibitem{Eckart1936}
C.~Eckart and G.~Young, ``The approximation of one matrix by another of lower
  rank,'' {\em Psychometrika}, vol.~1, 1936.

\bibitem{zahr2016phd}
M.~J. Zahr, {\em Adaptive Model Reduction to Accelerate Optimization Problems
  Governed by Partial Differential Equations}.
\newblock PhD thesis, Stanford University, August 2016.

\bibitem{Saad2003}
Y.~Saad, {\em Iterative Methods for Sparse Linear Systems, Second Edition}.
\newblock 2003.

\bibitem{An2008}
S.~S. An, T.~Kim, and D.~L. James, ``Optimizing cubature for efficient
  integration of subspace deformations,'' {\em ACM Transactions on Graphics},
  vol.~27, 2008.

\bibitem{Fang2013}
F.~Fang, C.~C. Pain, I.~M. Navon, A.~H. Elsheikh, J.~Du, and D.~Xiao,
  ``Non-linear petrov-galerkin methods for reduced order hyperbolic equations
  and discontinuous finite element methods,'' {\em Journal of Computational
  Physics}, vol.~234, 2013.

\bibitem{Huerta2003}
A.~H.~J. Donea, {\em Finite Element Methods for Flow Problems}.
\newblock 2003.

\end{thebibliography}
\appendix
\section{Singular Value Decomposition}
\label{app:Singular Value Decomposition}
This appendix provides details on the singular value decomposition (SVD) method used to find a basis matrix $\mathbf{\Phi}$ that satisfies the conditions presented in the main text. The procedure involves computing a truncated SVD \cite{Eckart1936} (TSVD) of the matrix $\mathbf{A}^u$, which is the matrix to be decomposed, with a given relative tolerance $0 < \epsilon_u \leq 1$. The TSVD function is defined symbolically as follows:

\begin{equation}
[\mathbf{U},\mathbf{\Sigma},\mathbf{V}] = \text{SVD}(\mathbf{A}^u,\epsilon_u),
\end{equation}

where $\mathbf{U}\in \mathbb{R}^{n\times r}$ is the matrix of left singular vectors, $\mathbf{\Sigma}\in \mathbb{R}^{r\times r}$ is the matrix of positive singular values (diagonal), and $\mathbf{V}\in\mathbb{R}^{m\times r}$ is the matrix of right singular vectors. These matrices satisfy the following relationships:

\begin{equation}
\mathbf{A}=\mathbf{U}\mathbf{\Sigma}\mathbf{V}^T+\mathbf{E},
\qquad
\|\mathbf{E}\|_F\leq\epsilon \|\mathbf{A}\|_F,
\qquad
\mathbf{U}^T\mathbf{U}=\mathbf{V}^T\mathbf{V}=\mathbf{I}_{r\times r},
\qquad
\Sigma_{(i+1,i+1)}\geq \Sigma_{(i,i)}>0,
\end{equation}
where $\|\cdot\|_F$ denotes the Frobenius norm, $\mathbf{E}$ is the truncation error, and $\Sigma{(i,i)}$ denotes the $i$-th largest singular value of $\mathbf{A}^u$. The basis matrix $\mathbf{\Phi}$ is then chosen as the matrix of the first $n$ columns of $\mathbf{U}$:
\begin{equation}
\mathbf{A}^u=\underbrace{\mathbf{U}_n}_{\mathbf{\Phi}}\mathbf{\Sigma}_n \mathbf{V}^T_n+\mathbf{E},
\end{equation}
where $\mathbf{U}_n$ and $\mathbf{V}_n$ are the first $n$ columns of $\mathbf{U}$ and $\mathbf{V}$, respectively, and $\mathbf{\Sigma}_n$ is the $n \times n$ diagonal matrix containing the $n$ largest singular values of $\mathbf{A}^u$.

\appendix
\end{document}